\documentclass[prl,twocolumn,superscriptaddress,a4paper,twoside]{revtex4}

\usepackage{amsmath}
\usepackage{amssymb}
\usepackage{todonotes}
\usepackage{nicefrac}
\usepackage{graphics}
\usepackage{color}
\usepackage{graphicx}
\makeatletter
\newcommand*{\rom}
[1]{\expandafter\@slowromancap\romannumeral #1@}
\makeatother


\begin{document}

\title{Effects of dynamical screening on the BCS-BEC crossover in double bilayer graphene: Density functional theory for exciton bilayers}

\author{F. Nilsson}
\email{fredrik.nilsson@teorfys.lu.se}
\author{F. Aryasetiawan}
\affiliation{Department of Physics, Division of Mathematical Physics, Lund University, Professorsgatan 1, 223 63, Lund, Sweden}

\begin{abstract}
We derive a gap equation for bilayer excitonic systems based on density functional theory and benchmark 
our results against quantum Monte-Carlo simulations and recent experiments on double bilayer graphene. The gap equation has a mean-field form but includes a consistent treatment of dynamical screening. We show that the gap survives at much higher densities than previously thought from mean-field estimates which gives strong indications that the double-bilayer graphene systems at zero magnetic field can be used as model systems to investigate the BCS-BEC crossover. Furthermore, we show that Josephson-like transfer of pairs can be substantial for small band gaps and densities.
\end{abstract}

\maketitle

In a recent experiment \cite{Burg2018} on two bilayer graphene sheets separated by a WSe$_2$ insulating barrier an enhanced tunneling rate was measured, which indicates the formation of an exciton condensate. While exciton formation has previously been achieved in similar systems in the quantum Hall regime at high magnetic fields \cite{liu2017,li2017} these were the first experimental indications of exciton formation in bilayer systems at zero magnetic field. In bilayer graphene systems the electron and hole concentrations are tunable by metallic gates, which opens up for the possibility to tune the device from the Bose-Einstein condensate (BEC) to the Bardeen-Cooper-Schrieffer (BCS) regime with possible applications in novel electronic devices \cite{Banerjee2009}. 

Theoretically, bilayer excitonic systems are typically described within mean-field theory \cite{lozovik2012,Perali2013,zarenia2014,Neilson2014,Conti2017,Conti2019} or using Quantum Monte-Carlo (QMC) simulations \cite{depalo2002,Maezono2013,rios2017}. While the latter is very accurate the calculations are restricted to simplified single-band systems which partly limit the predictive power. The mean-field calculations, on the other hand, can be applied to realistic systems but neglect dynamical screening as well as higher order self-energy diagrams and are therefore, as we will argue below, only accurate in the low-density limit. 
The role of dynamical screening for single bilayer graphene systems and topological insulators was previously investigated in Ref. \cite{sodemann2012} by solving the imaginary axis gap equation. Similar to the present work it was shown that a static mean-field description severely underestimates the gap in a large parameter regime. 
In this letter we derive a new gap equation, based on density functional theory,
which yields an effective account of dynamical screening at a sufficiently low computational cost to be applied to realistic materials. The method is applied to double bilayer graphene, which is the most promising bilayer candidate for achieving exciton condensation without an applied magnetic field. To our knowledge this is the first time that the effect of dynamical screening is studied for realistic multiband models of this material.
We begin by giving a brief account of the derivation (the complete derivation is given in the supplemental material (SM)), then we benchmark our method against QMC simulations for simplified models \cite{rios2017} and finally consider realistic systems where we compare to the experimental results in Ref. \cite{Burg2018} and mean-field calculations in Ref. \cite{Conti2019}. We show that, contrary to mean-field theories, the results from our effective gap equation qualitatively agree both with QMC simulations and experimental measurements. Due to its low computational cost the theory can provide an efficient means to predict exciton condensation in novel materials and an important complement to existing methods.  

We start with the Kohn-Sham Hamiltonian for the electron-hole system
\begin{align}
\hat{H}_{KS} = &\int dr \hat{\psi}^\dagger (r) H^e(r) \hat{\psi}(r) + \int dr \hat{\phi}^\dagger (r) H^h(r) \hat{\phi}(r)  \nonumber \\ 
&-\int dr dr' \Delta(r,r') \hat{\psi}^\dagger(r)\hat{\phi}^\dagger(r') + h.c.
\label{HamCorr}
\end{align}
Here $\psi(r)$ ($\phi(r)$) are the electron (hole) field operators and $\Delta(r,r')$ is the Kohn-Sham electron-hole potential.
Assuming band diagonal phase-coherent singlet coupling (i.e. electrons of spin $\sigma$, ,momentum $k$ and band $\gamma$ couple to holes with spin $-\sigma$ momentum $-k$ and band $\gamma$) the Hamiltonian can be rewritten as 
\begin{align}
 H_{KS} = &\sum_{k,\gamma} \varepsilon_{k\gamma}^h\hat{a}^\dagger_{k\gamma}\hat{a}_{k\gamma} + \sum_{k\gamma} \varepsilon_{k\gamma}^e\hat{b}^\dagger_{k\gamma}\hat{b}_{k\gamma}  \nonumber \\ -&\sum_{k\gamma}\Delta_{k\gamma}\hat{a}^{\dagger}_{k\gamma}\hat{b}^{\dagger}_{-k\gamma} + h.c.
 \label{HKS}
\end{align}
Here $\varepsilon_{k \gamma}^{e/h}$ are the eigenenergies of $H^{e/h}$, $\hat{b}^\dagger$ and $\hat{a}^\dagger$ are the corresponding creation operators and $\Delta_{k\gamma}$ is the matrix element of $\Delta(r,r')$ in the electron and hole eigenstates (see SM). 
This approximation is the analogue of the decoupling approximation in superconductor density functional theory (scDFT). 
For graphene bilayers in AB-stacking a tight-binding fit to the low-energy band structure is given by \cite{Partoens2006,mccann2013,Conti2019}
\begin{align}
\varepsilon_{k\gamma} = \frac{\gamma}{2}\sqrt{(t_1-\Gamma_k)^2 + \Omega_k}  - \mu  
\label{Eq:Ebilayer}
\end{align}
where
\begin{align*}
    &\Gamma_k=\sqrt{t_1^2 + (2\hbar v k)^2 + (2E_g\hbar v k)^2/t_1^2}, \\
    &\Omega_k= E_g^2 (1-(2\hbar vk)^2/t_1^2), \\
    &v=\sqrt{3} a t_0 / 2\hbar
\end{align*}
Unless otherwise specified we use the bilayer bandstructure above with the intercell distance $a=0.246$nm, intralayer hopping $t_0=3.16$eV and interlayer hopping $t_1=0.38$eV \cite{kuzmenko2009,Conti2019} for both the electron and hole layers. For simplicity we will assume $\varepsilon^e_{k\gamma}=\varepsilon^h_{-k\gamma} = \varepsilon_{k\gamma}$ in the derivation below, although it is straight-forward to generalize the derivation to different electron and hole dispersions. All calculations were performed at a temperature of 1.5 K.
The Hamiltonian in Eq. \ref{HKS} is diagonalized by a Bogoliubov transformation
\begin{align}
\hat{b}_{k\gamma}=u_{k\gamma}\hat{\alpha}_{k\gamma} + v_{k\gamma}\hat{\beta}^\dagger_{-k\gamma} \\
\hat{a}_{k\gamma}=u_{k\gamma}\hat{\beta}_{k\gamma} - v_{k\gamma}\hat{\alpha}^\dagger_{-k\gamma}.
\end{align}
The amplitudes $u_{k\gamma}$ and $v_{k\gamma}$ are subject to the equations
 \begin{align}
  u_{k\gamma}^2 = \frac{1}{2}(1 + \frac{\xi_{k\gamma}}{\tilde{E}_{k\gamma}}) \\
  v_{k\gamma}^2 = \frac{1}{2}(1 - \frac{\xi_{k\gamma}}{\tilde{E}_{k\gamma}}) \\
  2u_{k\gamma}v_{k\gamma} = \pm \Delta_{k\gamma}/E_{k\gamma},
 \end{align}
 where $\tilde{E}_{k\gamma}=\pm E_{k\gamma}$, $E_{k\gamma}=\sqrt{\varepsilon_{k\gamma}^2 + \Delta_{k\gamma}^2}$.
 We define the Kohn-Sham propagators
 \begin{align}
G^e(r\tau,r'\tau')&=-\langle \hat{T} \hat{\psi}(r\tau)\hat{\psi}^\dagger(r'\tau') \rangle_{KS}, \\
 G^h_{KS}(r\tau,r'\tau')&=-\langle \hat{T} \hat{\phi}^\dagger(r\tau)\hat{\phi}(r'\tau') \rangle_{KS},\\ 
 F^{eh}_{KS}(r\tau,r'\tau')&=-\langle \hat{T} \hat{\psi}(r\tau)\hat{\phi}(r'\tau') \rangle_{KS}.     
 \end{align}
  Utilizing the the decoupling approximation and Fourier transforming to frequency space yields
 
\begin{align}
 G^e_{k\gamma}(\omega_n)= 
 \frac{u_k^2}{i\omega_n - \varepsilon_{\gamma k}} + \frac{v_k^2}{i\omega_n + \varepsilon_{\gamma k}} \\
 G^h_{k\gamma}(\omega_n) =
 \frac{u_k^2}{i\omega_n + \varepsilon_{\gamma k}} + \frac{v_k^2}{i\omega_n - \varepsilon_{\gamma k}}\\
 F^{eh}_{k\gamma}(\omega_n)=
  \frac{v_k^{*}u_k}{i\omega_n - \epsilon_{\gamma k}} - \frac{u_kv_k^{*}}{i\omega_n + \epsilon_{\gamma k}},
\end{align}
Using the Sham-Schl\"uter connection \cite{Godby86Accurate,Godby87Quasi,Godby87Trends} with the same approximations as in scDFT 
\cite{Luders05Ab,ludersthesis,marquesthesis} one can then derive the gap equation for the Kohn-Sham system
 \begin{align}
 \Delta_{k\gamma} =   
 \frac{\sum_{\omega_n} \Sigma^{a}_{k\gamma}(\omega_n)G^h_{k\gamma}(\omega_n)G^e_{k\gamma}(\omega_n) ]}{\sum_{\omega_n} [G^h_{k\gamma}(\omega_n)G^e_{k\gamma}(\omega_n) ]} 
 \label{Eq:GapEq}
 \end{align}
 where $\Sigma^{a}$ is the exchange-correlation part of the anomalous self-energy and we have assumed that the effect of the normal self-energies are included in the bare dispersion. A generalized gap equation which include also the electron-electron and hole-hole correlations is given in the SM. It is straight-forward to show that a static exchange approximation to the self-energy yields the usual mean-field gap equation
 
 \begin{align}
  \Delta_{k\gamma} = -\sum_{k',\gamma'}\mathcal{F}_{kk'}^{\gamma\gamma'}\frac{V_{k-k'}\Delta_{\gamma'k'}}{2E_{\gamma'k'}}(1-f_\beta(\epsilon_{\gamma'k'+}) - f_\beta(\epsilon_{\gamma'k'-})).
  \label{Eq:GapEqMF}
 \end{align}
The geometrical form-factor $\mathcal{F}$ is related to the overlap of the single particle states $|\Psi_{k\gamma}\rangle \equiv |f_{k\gamma}\rangle e^{i\mathbf{k}\mathbf{r}}$ as $\mathcal{F}_{kk'}^{\gamma\gamma'} = \langle f_{k,\gamma} | f_{k',\gamma'} \rangle \langle f_{k',\gamma'} | f_{k,\gamma} \rangle$ \cite{lozovik2010ultra} (See the SM for details). 

In order to include the effect of dynamical screening we compute $\Sigma^{a}$ in Eq. \ref{Eq:GapEq} within the $GW$-approximation 
\begin{align}
\Sigma^{a}_{k\gamma}(i\omega_n) = \sum_{k',\gamma',\nu_l} \mathcal{F}_{k,k-k'}^{\gamma \gamma'}F^{eh}_{k-k',\gamma}(i\omega_n-i\nu_l)W^{eh}_{k'}(\nu_l)
\label{Eq:SE}
\end{align}
The screened electron-hole interaction for the bilayer system is given by \cite{lozovik2012,sodemann2012,Conti2019}
\begin{align}
W^{eh} = \frac{-V_D - \Pi_{a}(V^2 - V_D^2)}{1-2(V\Pi_{n} +V_D\Pi_{a}) +(\Pi_{n}^2 - \Pi_{a}^2)(V^2-V_D^2)}, 
\end{align}
where the $k$ and Matsubara indexes are implicit for all quantities and
\begin{align}
&V=V_k=\frac{2\pi e^2}{\epsilon}\frac{1}{|\mathbf{k}|} \\
&V_D=V^D_k = e^{-dk} \frac{2\pi e^2}{\epsilon}\frac{1}{|\mathbf{k}|}. 
\end{align}
Unless otherwise specified we use the dielectric constant $\epsilon=2$ for bilayer graphene sheets in a few layers of hexagonal boron nitride \cite{kumar2016,Conti2019}.
$\Pi_n=\Pi^{(n)}_q(i\nu_n)$ and $\Pi_a=\Pi^{(a)}_q(i\nu_n)$ are the normal and anomalous polarizations which are computed within the random-phase approximation (See SM).

In order to have an efficient implementation at low temperatures all Matsubara sums have to be done analytically. This is straightforward for the polarizations, but for the self-energy in Eq. \ref{Eq:SE} it is not as clear. In this work we adopt a similar strategy as Ref. \onlinecite{Akashi13development} and fit $W^{eh}_{k}(\nu_l)$ to a multiple plasmon pole model
\begin{align}
 \tilde{W}_{k}^{eh}(i\nu_m)=W_{k}^{eh}(0) + \sum_i^{N_p} a_{i;k}\left(\frac{2}{\omega_{i;k}}-\frac{2\omega_{i;k}}{\nu_m^2+\omega_{i;k}^2}\right).
\end{align}
We use up to 100 poles in order to get the correct shape of the peaks in Im$W^{eh}$. The position of the main pole and the relative weight of the poles are determined by the analytical continuation of $W^{eh}(i\nu_n)$ to the real axis, while the overall weight of the poles is determined by a least square fit of the imaginary axis data. We evaluated all Matsubara sums in Eqs. \ref{Eq:GapEq} and \ref{Eq:SE} analytically using the MatsubaraSum package for Mathematica \cite{MatsubaraSum}. The details of the fitting procedure are given in the SM.
\begin{figure*}[tb]
\begin{center}
\includegraphics[width=0.9\textwidth]{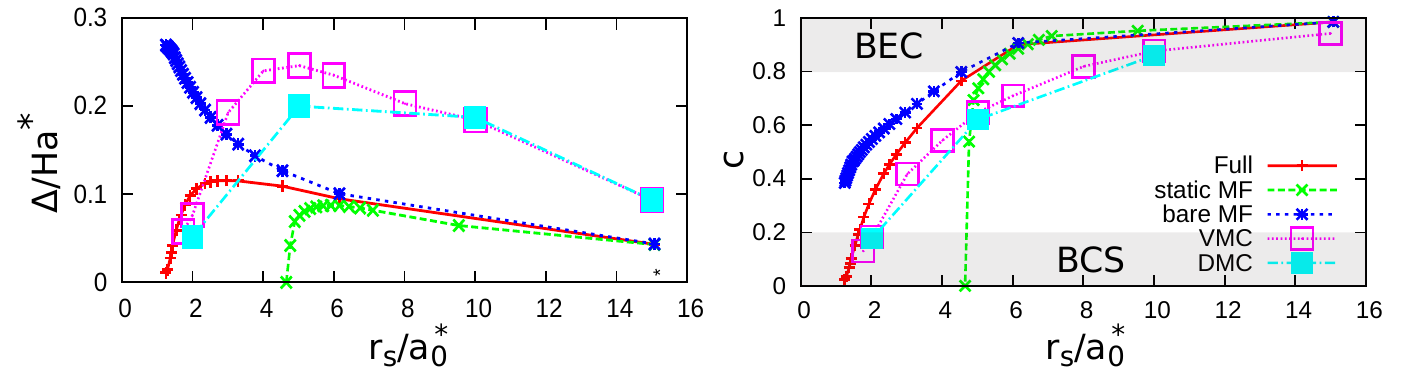} 
\caption{Maximum value of the superfluid gap (left) and condensate fraction (right) for a single band model with parabolic bands as a function of  $r_s/a_0^*$ where $r_s=\sqrt{1/(\pi n)}$ and $a_0^* = \kappa (m_e/m_e^*) a_0$. The variational quantum Monte Carlo (VMC) and diffusion Quantum Monte-Carlo (DMC) data are taken from Ref. \onlinecite{rios2017}.}
\label{fig:compQMC}%
\end{center}
\end{figure*}

To benchmark our method we first consider a model with single parabolic electron and hole bands and compare with the Quantum Monte-Carlo simulations by Rìos \emph{et. al} \cite{rios2017}. 
In Fig. \ref{fig:compQMC} we compare the results from solving the full DFT gap equation (Eq. \ref{Eq:GapEq}), the standard mean-field gap equation (Eq. \ref{Eq:GapEqMF}) with the static value of the screened interaction (screened MF-equation) and the same equation but with the unscreened value of the interaction (bare MF-equation) with the corresponding results from the QMC simulations in Ref. \onlinecite{rios2017}. The screened MF equation was benchmarked against QMC previously, with encouraging results \cite{Neilson2014}. However, in these calculations a different form of the screened interaction was used and the formfactor was ignored in the gap equation. As we will see below, a correct treatment of the interaction and formfactor yields a larger screening and substantially worsens the performance of the screened MF equation.
We begin by considering the maximum value of the gap. To facilitate the comparison we adopt the scaled units of Ref. \onlinecite{rios2017} with the effective energy unit Ha$^*=m_e^*/m_e \kappa^{-2}$Ha and distances in units of $a_0^* = \kappa (m_e/m_e^*) a_0$. The density $n$ is defined through $r_s/a_0^*$ where $r_s=\sqrt{1/(\pi n)}$. We use the same interlayer distance $d/a_0^* = 0.4$.
In Ref. \onlinecite{rios2017} the gap was estimated by fitting the excitation energies $E(k)$ obtained from the QMC simulations to an effective mean-field gap-equation with $k$-independent gap 
\begin{align*}
    E(k) = \sqrt{(k^2/2m^* - \mu)^2 + \Delta^2}
\end{align*}
with $m^*$, $\mu$ and $\Delta$ as fitting variables. In our simulations we use the effective mass which gave the optimal fit for each density (extracted from Fig. 3 in Ref \onlinecite{rios2017}) and consider the maximum value of the $k$-dependent gap.  

 At low densities (high $r_s/a_0^*$) all mean-field calculations yield the same gap (left panel in Fig.~\ref{fig:compQMC}). As the density is increased the screening reduces the gap of the screened MF result compared to the bare MF results substantially and at $r_s/a_0^*\approx 4.5$ the screened mean-field gap rapidly goes to zero. The additional correction in the DFT gap equation reduces the screening and increases the gap compared to the static MF results, while it is always smaller than the corresponding value in the bare MF calculation. Hence the DFT gap equation incorporates the frequency dependence of the interaction in an effective static theory, similar to the $Z$-factor approach for Hubbard models \cite{casula12b}.
Due to the decrease of the effective screening the full gap goes to zero around $r_s/a_0^* \approx 1.2-2$, which corresponds to much higher densities than the static MF gap and is in qualitative agreement with the QMC results. At low and intermediate densities ($r_s/a_0^* > 4$) all mean-field calculations underestimate the gap compared to the QMC data. Since the bare MF calculations sets the upper limit of $\Delta$ obtainable by any mean-field approximation relying on the first-order diagram of $\Sigma$, this indicates that the differences at low density may originate from many-body effects beyond a first-order mean-field description. Both the screened mean-field and DFT gap equation are restricted to the first-order term of the self-energy expansion in $W^{eh}$. This is a good approximation as long as $W^{eh}$ is sufficiently small or equivalently the screening is large,  which explains why the DFT gap-equation agrees better with the QMC data for high densities than for intermediate and low densities. Thus, the poor performance of the screened MF equation for high densities is related to the neglect of the frequency dependence of the interaction, which is to a large extent cured by the effective treatment in the DFT gap-equation. The mean-field treatments give reasonable results also in the low-density limit, which is expected since the mean-field treatment becomes exact in the limit $n\rightarrow 0$. It should be noted that the DFT gap-equation (Eq. \ref{Eq:GapEq}) is not restricted to the $GW$-approximation for the self-energy, and hence in principle higher order self-energy terms can be accounted for within our framework. 

\begin{figure*}[tb]
\begin{center}
\includegraphics[width=0.9\textwidth]{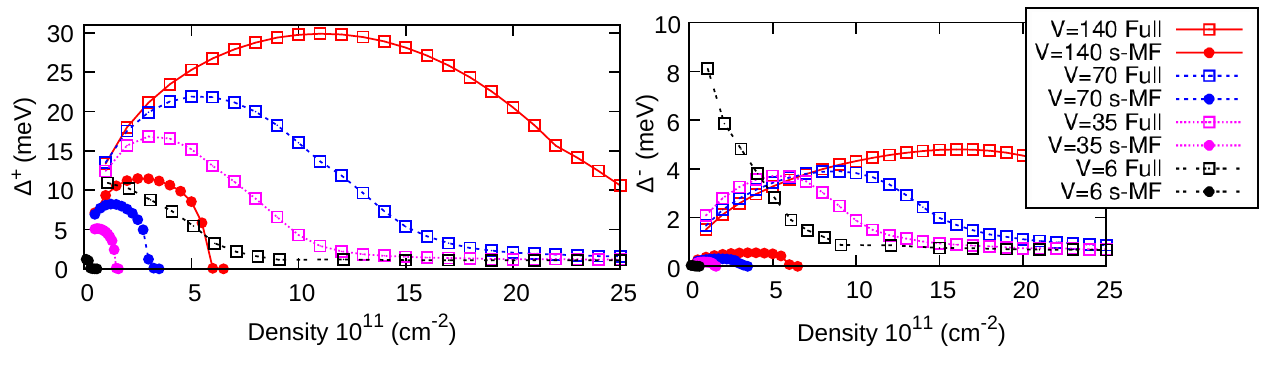} 
\caption{Maximum superfluid gap for the conduction (left) and valence band (right) as a function of the electron and hole density for different band gaps (140meV, 70 meV, 35meV and 6meV). The results from solving the full gap equation are compared to the static mean-field results. We used the bare electron and hole dispersion in Eq. \ref{Eq:Ebilayer} with the parameters in the text.}
\label{fig:compPRB}%
\end{center}
\end{figure*}

In the right panel of Fig \ref{fig:compQMC} we compare the condensate fractions $c$. In our mean-field results the condensate fractions can be estimated from the Bogoliubov amplitudes as 
\begin{align}
c = \frac{\sum_k u_k^2 v_k^2}{\sum_k v_k^2},
\label{Eq:confrac}
\end{align}
while in the QMC calculations it was estimated by interpolating the two-body rotationally averaged density matrix. The transition to the BCS phase is characterized by a value of $c<0.2$. 
Contrary to the QMC calculations the static MF simulations predicts $c$ to vanish at $r_s/a_0^*\approx 4.5$, before the system has entered the BCS phase.
The bare MF calculation on the other hand predicts a too high value of $c$ for small $r_s/a_0^*$. 

\begin{figure}[tb]
\begin{center}
\includegraphics[width=0.45\textwidth]{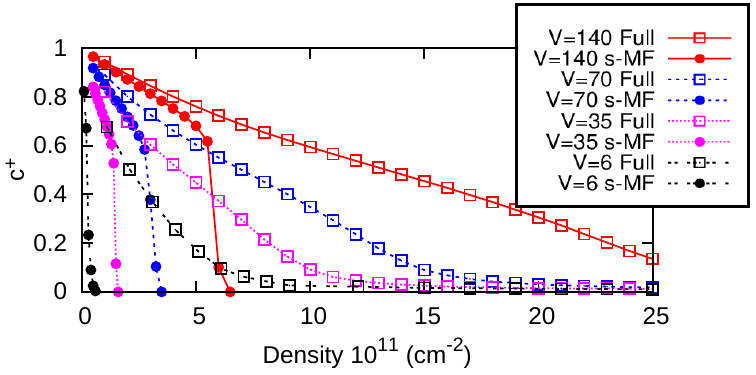} 
\caption{Condensate fraction for the conduction band as a function of the electron and hole density for the same systems as in Fig. \ref{fig:compPRB}.}
\label{fig:condfrac}%
\end{center}
\end{figure}

The full calculations agree qualitatively well with the QMC results, especially for small values of $r_s/a_0^*$ where both calculations predicts a transition to the BCS phase at around $r_s/a_0^* \approx 2$. For low and intermediate densities (intermediate-high $r_s/a_0^*$) the full gap equation overestimates $c$ compared to the QMC data and predicts the crossover to the BEC phase with $c\approx 0.8$ at $r_s/a_0^* \approx 4$, which can be compared to a transition at  $r_s/a_0^* \approx 7$ in the QMC simulations. The overestimation of $c$ at intermediate and low densities is inherent to all mean-field approximations and is again related to the neglect of higher order self-energy terms. However, the inclusion of the dynamical screening in the full gap equation improves the static MF results substantially and yields results which at least qualitatively agree with the QMC simulations in Ref. \onlinecite{rios2017}. It should be noted that in more realistic multiband models the screening is larger which, according to the discussion above, indicates that the DFT gap equation is accurate for a larger density range for these systems than for the single-band model discussed above. 
So far we have benchmarked our method against QMC simulations for simple single-band models.
In the remainder of this letter we will focus on more realistic models for double bilayer graphene with the single-particle dispersion relation in Eq. \ref{Eq:Ebilayer}. These systems were extensively studied in Ref. \onlinecite{Conti2019} within a static mean-field approximation for band gaps $\leq$ 210 meV, which is close to the maximum gap obtainable for these systems ($\approx$ 300 meV) \cite{Ohta2006,Hongki2007,Castro2007,oostinga2008,mccann2013}. In Fig. \ref{fig:compPRB} we show the superfluid gaps ($\Delta^\pm$) for the conduction band ($\gamma=+1$) and valence band ($\gamma=-1$) channels. The strength of the interband screening depends intrinsically on the bandgap and the larger the band gap the smaller the screening. Therefore $\Delta$ is larger and survives at higher densities for large band gaps. However, for large band gaps the condensate fraction (Fig. \ref{fig:condfrac}) also decreases more slowly as  function of the density. Due to these counteracting tendencies the static mean-field approximation predicts that $\Delta$ vanishes in the crossover region with $0.2<c^+<0.8$ and that no system, independent of the band gap, reaches the BCS state at $c^+\approx 0.2$. The full calculations, based on the DFT gap equation, yield a fundamentally different conclusion. In these calculations the effect of screening is reduced substantially which yields larger superfluid gaps that survive at higher densities. The picture of a relatively constant $\Delta$ that rapidly goes to zero at a critical density where the screening sets in is replaced by a slowly decaying gap for high densities. 
Looking at $c^+$ in Fig. \ref{fig:condfrac} one can see that the BCS state is reached before the gap vanishes, \emph{for all band gaps considered}. Furthermore, the slowly decaying tail of the gap function and the condensate fractions in Figs. \ref{fig:compPRB}-\ref{fig:condfrac} suggest that, for the ideal system, it may be possible to stabilize a condensate with a small gap ($\approx$1 meV) and a small condensate fraction ($\approx$ 1$\%$) for high densities also in systems with small band gaps.   

Our static mean-field calculations agree with Ref. \cite{Conti2019}, and similar to these calculations $\Delta^+ >> \Delta^-$ for all bandgaps and densities, which suggest that the two gap equations are decoupled and that a Josephson-like transfer of pairs is negligible at all densities. In the full calculations the reduced screening yields much larger gaps. For small bandgaps $\Delta \approx E_g$, which yields a strong coupling of the gap equations and $\Delta^+ \approx \Delta^-$. Hence, contrary to the MF calculations we predict that Josephson-like transfer of pairs is substantial for small bandgaps.

In Ref.~\onlinecite{Burg2018} a double bilayer graphene system with WSe$_2$ dielectric barrier of 1.4nm thickness was investigated experimentally. The carrier populations and effective band-gap were controlled by external gates and the tunneling current ($I_{\text{int}}$) was measured. Around $n_e=n_h=7 \cdot 10^{11}$ cm$^{-2}$  the differential tunneling conductance $g_{\text{int}} = dI_\text{int}/dV_{int}$ was strongly enhanced, which cannot be explained with a single-band tunneling model and hence signals the formation of an interbilayer exciton condensate. The enhanced tunneling was measured at gate voltages corresponding to a bandgap of 6 meV in the single-band model, although the actual bandgap can vary slightly from this value due to many-body effects \footnote{Private communication with N. Prasad.}. The system we consider in Fig. \ref{fig:compPRB} has a thinner barrier and an hBN dielectric, which has a smaller dielectric constant than the WSe$_2$ barrier in the experimental setup. Hence the calculations in this paper can be considered to provide an upper limit for the superfluid gap in the experimental setup. Within the static mean-field description the gap vanishes already around the density  $0.3 \cdot 10^{11}$ cm$^{-2}$ for a bandgap of 6 meV, and therefore these calculations qualitatively fails to explain the experimental observations. The full calculation on the other hand has a nonvanishing gap at $n=7 \cdot 10^{11}$ cm$^{-2}$, although the maximum gap is at lower densities.

To summarize we have derived a new gap equation for bilayer excitonic systems which includes an effective treatment of the frequency dependence of the electron-hole interaction. The frequency dependence has a dramatic effect on the superfluid gap, which is enhanced substantially and survives at much higher densities. Contrary to the usual static mean-field treatment our results qualitatively agree with both to QMC simulations for simplified systems \cite{rios2017} and experimental measurements for real systems \cite{Burg2018}. Our method therefore provides an important complement to existing methods to describe bilayer excitonic systems. For double bilayer graphene at zero magnetic field our calculations predict non-vanishing superfluid gaps across the entire BEC-BCS crossover. Double bilayer graphene may therefore provide a  convenient solid-state systems that can be used to investigate the new emerging physics in the crossover regime as well as to design new devices. Contrary to static mean-field calculations we predict that the superfluid gaps are substantial even for small band gaps ($<10$ meV) and that Josephson-like transfer of pairs can be substantial. 

\begin{acknowledgments}
F.N. and F.A. acknowledge financial support from the Knut and Alice Wallenberg Foundation and the Swedish Research Council (Vetenskapsrådet, VR).
The computations were performed on resources provided by the Swedish National Infrastructure for Computing (SNIC) at LUNARC. F. N. would like to thank N. Prasad and W. Burg for helpful discussions regarding the experimental results in Ref. \cite{Burg2018}
\end{acknowledgments}

\bibliography{refs}

\end{document}


\title{Supplemental information for "Effects of dynamical screening on the BCS-BEC crossover in double bilayer graphene: Density functional theory for exciton bilayers"}
\author{F. Nilsson}
\email{fredrik.nilsson@teorfys.lu.se}
\affiliation{Department of Physics, Division of Mathematical Physics, Lund University, Professorsgatan 1, 223 63 Lund, Sweden}
\author{F. Aryasetiawan}
\affiliation{Department of Physics, Division of Mathematical Physics, Lund University, Professorsgatan 1, 223 63 Lund, Sweden}

\maketitle

\section{Frequency dependence of the screened interaction}
\begin{figure}[tb]
\begin{center}
\includegraphics[width=0.45\textwidth]{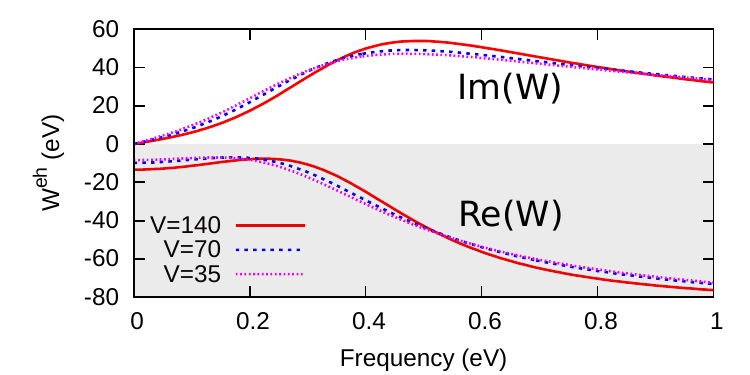} 
\caption{Screened electron-hole interaction from the self-consistent solution of the full gap-equation (Eq. 15 of the main text) at $k=k_F$ at the density $n=5\cdot 10^{11}$ cm$^{-2}$ for the bandgaps 140, 70 and 35 meV.}
\label{fig:W}%
\end{center}
\end{figure}

To understand why the frequency dependence of the interaction has such a dramatic effect on the gap we consider $W(\omega)$, obtained by Pad\'e analytic continuation, in Fig \ref{fig:W} at $k=k_F$ and $n=5\cdot 10^{11}$ cm$^{-2}$ for different bandgaps. A static approximation to $W$ is sufficient if Re$W(\omega)$ has a weak frequency dependence for low frequencies. From Fig. \ref{fig:W} it is clear that this is not the case for these systems. On the contrary, the pole in Im$W$ occurs at low frequencies around 500meV  which yields a corresponding Kramer's Kronig feature in the real part. Therefore these systems are not well described by a purely static approximation to $W$.  

\section{Derivation of gap equation}
\subsection{Density Functional Theory}
DFT for excitonic bilayers can be derived in close anology to DFT for superconductors (scDFT) \cite{Luders05Ab,ludersthesis,marquesthesis}
The electron-hole Hamiltonian of the bilayer system can in second-quantized form be expressed as
\begin{widetext}
\begin{align}
\hat{H} &= \overbrace{\int_R dr \, \hat{\psi}^\dagger(r) \left(- \frac{\nabla^2}{2} + V^{e}_{crys}(\mathbf{r})  \right) \hat{\psi}(r)}^{\hat{H}_0^e} + \overbrace{\int_L dr \, \hat{\phi}^\dagger(r) \left( \frac{\nabla^2}{2} + V^{h}_{crys}(\mathbf{r}) \right) \hat{\phi}(r)}^{\hat{H}_0^h}   \nonumber \\
  &+ \overbrace{\frac{1}{2}\int_R drdr' \hat{\psi}^\dagger(r) \hat{\psi}^\dagger(r')  \frac{1}{|\mathbf{r}-\mathbf{r}'|}  \hat{\psi}(r')\hat{\psi}(r)}^{\hat{U}^{ee}}
  + \overbrace{\frac{1}{2}\int_L drdr' \hat{\phi}^\dagger(r) \hat{\phi}^\dagger(r')  \frac{1}{|\mathbf{r}-\mathbf{r}'|}  \hat{\phi}(r')\hat{\phi}(r)}^{\hat{U}^{hh}} \nonumber \\
    &+ \underbrace{\frac{1}{2}\int_R dr\int_L dr' \hat{\psi}^\dagger(r) \hat{\phi}(r')  \frac{1}{|\mathbf{r}-\mathbf{r}'|}  \hat{\phi}^\dagger(r')\hat{\psi}(r)
     + \frac{1}{2}\int_L dr\int_R dr' \hat{\phi}(r) \hat{\psi}^\dagger(r')  \frac{1}{|\mathbf{r}-\mathbf{r}'|}  \hat{\psi}(r')\hat{\phi}^\dagger(r)}_{^{\hat{U}^{eh}}}
     \label{Hfull}
\end{align}
\end{widetext}
where $\hat{\psi(r)}$ ($\hat{\phi(r)}$) are the electron (hole) field operators and all terms that can be incorporated as a shift of the chemical potential have been ignored. The electrons and holes are assumed to be confined on their respective layers and electron-hole recombination is ignored. The generalized coordinate $r$ includes both the spatial $\mathbf{r}$ and spin $\sigma$ degrees of freedom $r=(\mathbf{r},\sigma)$.
Now we want to map this Hamiltonian to a noninteracting Kohn-Sham system. 
To formulate a Hohenberg-Kohn like statement we have to generalize the Hamiltonian
in Eq. \ref{Hfull} to
\begin{align}
\hat{H}= \hat{T}^e + \hat{T}^h + \hat{U}^{ee}+ \hat{U}^{hh}+ \hat{U}^{eh} + \hat{V}^e_{crys} +\hat{V}^h_{crys} + \hat{\Delta} - \mu^e \hat{N}^e- \mu^h \hat{N}^h
\end{align}
where
\begin{align}
\hat{V}^{e}_{crys} &= \sum_\sigma \int_R d^3r\hat{\psi}^\dagger(\mathbf{r},\sigma) V^e_{crys}(\mathbf{r}) \hat{\psi}(\mathbf{r},\sigma), \\
\hat{V}^{h}_{crys} &= \sum_\sigma \int_R d^3r\hat{\phi}^\dagger(\mathbf{r},\sigma) V^h_{crys}(\mathbf{r}) \hat{\phi}(\mathbf{r},\sigma),
\end{align}

\begin{align}
\hat{\Delta} = \sum_{\sigma \sigma'} \int d^3r d^3 r' \left[ \hat{\psi}^\dagger (\mathbf{r},\sigma) \Delta(\mathbf{r},\mathbf{r}') \hat{\phi}^\dagger(\mathbf{r}',\sigma') + h.c. \right],
\end{align}
and $\hat{T}^e$ and $\hat{T}^h$ are the electron and hole kinetic energy operators, respectively. 
In analogy with Ref. \onlinecite{Luders05Ab} this formulation is a multicomponent formulation of DFT that relies on three densities,
\begin{enumerate}
 \item The electron density $n^e(\mathbf{r})= \sum_{\sigma} \langle \hat{\psi}^\dagger(\mathbf{r},\sigma) \hat{\psi}(\mathbf{r},\sigma)\rangle$.
  \item The hole density $n^h(\mathbf{r})= \sum_{\sigma} \langle \hat{\phi}^\dagger(\mathbf{r},\sigma) \hat{\phi}(\mathbf{r},\sigma)\rangle$.
 \item The exciton density $\chi(\mathbf{r},\mathbf{r}') = \sum_{\sigma,\sigma'} \langle \hat{\psi}^\dagger(\mathbf{r},\sigma) \hat{\phi}^\dagger(\mathbf{r}',\sigma')\rangle$,
 which also is the order parameter characterizing the exciton formation.
\end{enumerate}

According to the Hohenberg-Kohn theorem there is, in thermal equilibrium, a one-to-one mapping between the densities $n^e(\mathbf{r})$,$n^h(\mathbf{r})$, $\chi(\mathbf{r},\mathbf{r}')$ and their conjugate 
potentials $V^e(\mathbf{r})$, $V^h(\mathbf{r})$, and $\Delta(\mathbf{r},\mathbf{r}')$.
The grand canonical potential $\Omega$ is minimized by the equilibrium densities
\begin{align}
 &\Omega[n^e,n^h,\chi] = F[n,\chi] + \int_R d^3 r n^e(\mathbf{r})[V^e_{crys}(\mathbf{r})-\mu^e] + \nonumber \\ 
 &\int_L d^3 r n^h(\mathbf{r})[V^h_{crys}(\mathbf{r}) 
 -\mu^h] - \int d^3 r d^3 r' [\chi(\mathbf{r},\mathbf{r}')\Delta(\mathbf{r},\mathbf{r}') + h.c.]
\end{align}
where the universal function $F$ is defined as
\begin{align}
 &F[n,\chi] = T^e[n^e,n^h,\chi]+ T^h[n^e,n^h,\chi] + U^{ee}[n^e,n^h,\chi] \nonumber \\ 
 &+ U^{hh}[n^e,n^h,\chi]+ U^{eh}[n^e,,n^h,\chi]- \frac{1}{\beta} S[n,\chi],
\end{align}
$S$ is the entropy\cite{Mermin65Thermal}
\begin{align}
S[n,\chi] = -\mathrm{Tr}\{ \hat{\rho}_0[n,\chi ] \}
\end{align}
and $\hat{\rho}_0$ the grand canonical density matrix 
\begin{align}
 \hat{\rho}_0 = e^{-\beta (\hat{H}-\mu^e\hat{N}^e-\mu^h\hat{N}^h)}/\mathrm{Tr}e^{-\beta (\hat{H}-\mu^e\hat{N}^e-\mu^h\hat{N}^h)}.
\end{align}

Now we want to map this to a non-interacting Kohn-Sham system with the same ground-state density as the interacting system.
The Kohn-Sham system is described by the following thermodynamic potential:
\begin{align}
 &\Omega_{KS}[n^e,n^h,\chi] = F_{KS}[n^e,n^h,\chi] + \int_R d^3 r n^e(\mathbf{r})[v^e_{KS}(\mathbf{r})-\mu^e] \nonumber  \\
 &+\int_L d^3 r n^h(\mathbf{r})[v^h_{KS}(\mathbf{r})-\mu^h] - \int d^3 r d^3 r' [\chi(\mathbf{r},\mathbf{r}')\Delta_{KS}(\mathbf{r},\mathbf{r}') + h.c.].
\end{align}
Here we have defined
\begin{align}
 F_{KS}[n^e,n^h,\chi] = &T_{KS}^e[n^e,n^h,\chi] + T_{KS}^h[n^e,n^h,\chi]  \nonumber \\
 &- \frac{1}{\beta} S_{KS}[n^e,n^h,\chi], 
\end{align}
\begin{align}
v^{e/h}_{KS}[n^e,n^h,\chi](\mathbf{r}) = V^{e/h}_{crys}(\mathbf{r}) + v^{e/h}_{xc}[n^e,n^h,\chi](\mathbf{r}) + \int_R d^3 r' \frac{n^{e/h}(\mathbf{r}')}{|\mathbf{r}-\mathbf{r}'|}
\end{align}
and
\begin{align}
 \Delta_{KS}[n^e,n^h,\chi](\mathbf{r},\mathbf{r}') = - \frac{\chi(\mathbf{r},\mathbf{r}')}{|\mathbf{r}-\mathbf{r}'|} + \Delta_{xc}[n^e,n^h,\chi](\mathbf{r},\mathbf{r}').
\end{align}

The exchange-correlation potentials are defined as functional derivatives of the exchange correlation free energy functional
\begin{align}
 v_{xc}^{e/h}[n^e,n^h,\chi](\mathbf{r}) = \frac{\delta F_{xc}[n^e,n^h,\chi]}{\delta n^{e/h}(\mathbf{r})}
\end{align}
\begin{align}
 \Delta_{xc}[n^e,n^h,\chi](\mathbf{r}\mathbf{r}') = \frac{\delta F_{xc}[n^e,n^h,\chi]}{\delta \chi(\mathbf{r},\mathbf{r}')}
\end{align}
and $F_{xc}$ is defined through the free energy functional
\begin{align}
F[n^e,n^h,\chi] = F_{KS}[n^e,n^h,\chi] + F_{xc}[n^e,n^h,\chi] + E_{H}[n^e,n^h,\chi] .
\end{align}
 
 The Hartree energy contains three contributions
\begin{align}
 &E_H[n^e,n^h,\chi] = \frac{1}{2} \int_R d^3r d^3r' \frac{n^e(\mathbf{r})n^e(\mathbf{r}')}{|\mathbf{r}-\mathbf{r}'|} 
 \nonumber \\ &+\frac{1}{2} \int_L d^3r d^3r' \frac{n^h(\mathbf{r})n^h(\mathbf{r}')}{|\mathbf{r}-\mathbf{r}'|}
 + \int d^3r d^3r' \frac{|\chi(\mathbf{r},\mathbf{r}')|^2}{|\mathbf{r}-\mathbf{r}'|} .
\end{align}

\subsection{Bogoliubov equations}
The Kohn-Sham Hamiltonian is given by
\begin{align}
\hat{H}_{KS} = &\int dr \hat{\psi}^\dagger (r) H^e \hat{\psi}(r) + \int dr \hat{\phi}^\dagger (r) H^h \hat{\phi}(r)  \nonumber \\ 
&-\int dr dr' \Delta_{KS}(r,r') \hat{\psi}^\dagger(r)\hat{\phi}^\dagger(r') + h.c.
\label{HamCorr}
\end{align}
where $H^{e/h} = \mp \nabla^2_{e/h}/2 + v_{KS}^{e/h}$.
Following Refs. \onlinecite{Luders05Ab} and \onlinecite{gennes1966}  we then define
\begin{align}
 \hat{\psi}(r) &= \sum_m [u_m^e(r)\hat{\gamma}_{1m} - v_m^{e*}(r)\hat{\gamma}_{2m}^\dagger] \\
 \hat{\phi}(r) &= \sum_m [u_m^h(r)\hat{\gamma}_{2m} + v_m^{h*}(r)\hat{\gamma}_{1m}^\dagger] 
\end{align}
where the operators $\gamma$ follow Fermionic anti-commutation relations:
\begin{align}
&\hat{\gamma}^\dagger_{\alpha n}\hat{\gamma}_{\beta m} + \hat{\gamma}_{\beta m} \hat{\gamma}^\dagger_{\alpha n} = \delta_{mn}\delta_{\alpha \beta} \\
&\hat{\gamma}_{\alpha n}\hat{\gamma}_{\beta m} + \hat{\gamma}_{\beta m} \hat{\gamma}_{\alpha n} = 0
\end{align}
Furthermore the transformation is chosen to diagonalize $H_{KS}$
\begin{align}
\hat{H}_{KS} = E_g + \sum_{m,\alpha} \varepsilon_{\alpha m}\hat{\gamma}^\dagger_{\alpha m}\hat{\gamma}_{\alpha m}
\end{align}
where $E_g$ is the ground state energy. This implies that the commutators between the Hamiltonian and the operators are given by:
\begin{align}
 [\hat{H}_{KS},\hat{\gamma}_{\alpha m}]=&-\varepsilon_{\alpha m}\hat{\gamma}_{\alpha m} \\
  [\hat{H}_{KS},\hat{\gamma}^\dagger_{\alpha m}]=&\varepsilon_{\alpha m}\hat{\gamma}^\dagger_{\alpha m}
\end{align}

From the Fermionic anti-commutation relations of the electron and hole field operators the commutator between the hole and electron annihilation operators and the Hamiltonian are given by:
\begin{align}
[\hat{\psi}(r),\hat{H}_{KS}] &= H_e(r)\hat{\psi}(r) + \int dr' \Delta(r,r') \hat{\phi}^\dagger(r') \\
[\hat{\phi}(r),\hat{H}_{KS}] &= H_h(r)\hat{\phi}(r) - \int dr' \Delta(r,r') \hat{\psi}^\dagger(r')
\end{align}

Now inserting the expressions for the operators in terms of the $\hat{\gamma}$:s, making use of the commutation relation for the $\hat{\gamma}$:s on the left hand side and equating equal coefficients of the $\hat{\gamma}$:s 
yields the Kohn-Sham Bogoliubov-de Gennes (KS-BdG) equations:

\begin{align}
 H_e(r)u_m^e(r) + \int dr' \Delta(r,r') v_m^h(r') =& u_m^e(r)\varepsilon_{1m} \nonumber \\
 -H^*_e(r)v_m^e(r) + \int dr' \Delta(r,r')^* u_m^h(r') =& v_m^e(r)\varepsilon_{2m} \nonumber \\
 H_h(r)u_m^h(r) + \int dr' \Delta(r,r') v_m^e(r') =& u_m^h(r)\varepsilon_{2m} \nonumber \\
-H^*_h(r)v_m^h(r) + \int dr' \Delta(r,r')^* u_m^e(r') =& v_m^h(r)\varepsilon_{1m}.
\label{KSB-final2}
\end{align}

\subsection{Decoupling approximation}
Since the electron and hole states are solutions to separate Hamiltonians we need to specify
which states that should couple in the decoupling approximation. In practice we will consider band-diagonal singlet phase-coherent coupling,
i.e.  electrons from band $\gamma$ with momentum $k$ and spin $\sigma$ couple to holes in band $\gamma$ with momentum $-k$ and spin $-\sigma$. 
This is similar to Ref \onlinecite{lozovik2012} which is the appropriate choice for $s$-wave band diagonal pairing. This yields
\begin{align}
u_{\mathbf{k}\gamma}^{e/h}(\mathbf{r}) \approx u_{\mathbf{k}\gamma} \phi_{\pm \mathbf{k} \gamma}^{e/h}(\mathbf{r}) \\
v_{\mathbf{k}\gamma}^{e/h}(\mathbf{r}) \approx v_{\mathbf{k}\gamma} \phi_{\pm \mathbf{k} \gamma}^{e/h}(\mathbf{r}) \\
H_{e/h}\phi_{\mathbf{k}\gamma}^{e/h}(\mathbf{r}) = E^{e/h}_{\mathbf{k}\gamma} \phi_{\mathbf{k}\gamma}^{e/h}(\mathbf{r}) 
\end{align}
where we have assumed a paramagnetic electron and hole bandstructure, i.e. $\phi_{\mathbf{k}\gamma}^{e/h}(\mathbf{r}\uparrow)=\phi_{\mathbf{k}\gamma}^{e/h}(\mathbf{r}\downarrow)=\phi_{\mathbf{k}\gamma}^{e/h}(\mathbf{r})$.
Inserting these expressions in the Bougoliubov equations and using the orthornormality of the Kohn-Sham orbitals yields the following equations for the eigenvalues and coefficients $u_{\mathbf{k}\gamma}$ and $v_{\mathbf{k}\gamma}$:
\begin{gather}
 \begin{bmatrix} E_{\mathbf{k}\gamma}^e & \Delta_{\mathbf{k}\gamma} \\ \Delta^*_{\mathbf{k}\gamma} & -E_{-\mathbf{k} \gamma}^h \end{bmatrix}
 \begin{bmatrix} u_{\mathbf{k}\gamma} \\ v_{\mathbf{k}\gamma} \end{bmatrix}
 = \varepsilon_{1\mathbf{k}\gamma}
  \begin{bmatrix} u_{\mathbf{k}\gamma} \\ v_{\mathbf{k}\gamma} \end{bmatrix},
  \label{Eq:Eigenbog1}
 \end{gather}
\begin{gather}
 \begin{bmatrix} E_{-\mathbf{k} \gamma}^h & \Delta_{\mathbf{k}\gamma} \\ \Delta^*_{\mathbf{k}\gamma} & -E_{\mathbf{k}\gamma}^e \end{bmatrix}
 \begin{bmatrix} u_{\mathbf{k}\gamma} \\ v_{\mathbf{k}\gamma} \end{bmatrix}
 = \varepsilon_{2\mathbf{k}\gamma}
  \begin{bmatrix} u_{\mathbf{k}\gamma} \\ v_{\mathbf{k}\gamma} \end{bmatrix}
  \label{Eq:Eigenbog2}
 \end{gather}
where 
\begin{align}
     \Delta_{\mathbf{k}\gamma} &= \int d^3r d^3r' \phi^{e*}_{\mathbf{k}\gamma}(\mathbf{r})\Delta(\mathbf{r},\mathbf{r}')\phi^h_{-\mathbf{k}\gamma}(\mathbf{r}').
\end{align}
 The solutions of Eqs. \ref{Eq:Eigenbog1}-\ref{Eq:Eigenbog2} are
\begin{align}
 \varepsilon_{1 \mathbf{k}\gamma} &= \delta_{\mathbf{k}\gamma} + \tilde{E}_{\mathbf{k}\gamma} \\
 \varepsilon_{2 \mathbf{k}\gamma} &= 
 -\delta_{\mathbf{k}\gamma} + \tilde{E}_{\mathbf{k}\gamma} \\
 \end{align}
 where we have defined
 $E_{\mathbf{k}\gamma} = \sqrt{\xi_{\mathbf{k}\gamma}^2 + \Delta_{\mathbf{k}\gamma}^2}$,
 $\xi_{\mathbf{k}\gamma} = (E_{\mathbf{k}\gamma}^e + E_{-\mathbf{k}\gamma}^h)/2$, $\delta_{\mathbf{k}\gamma}=\frac{E^e_{-\mathbf{k}\gamma}-E^h_{\mathbf{k}\gamma}}{2}$, $\tilde{E}_{\mathbf{k} \gamma} = \pm E_{\mathbf{k} \gamma}$.

 The corresponding amplitudes $u_{\mathbf{k}\gamma}$ and $v_{\mathbf{k}\gamma}$ are subject to the equations
 \begin{align}
  u_{\mathbf{k}\gamma}^2 = \frac{1}{2}(1 + \frac{\xi_{\mathbf{k}\gamma}}{\tilde{E}_{\mathbf{k}\gamma}}), \\
  v_{\mathbf{k}\gamma}^2 = \frac{1}{2}(1 - \frac{\xi_{\mathbf{k}\gamma}}{\tilde{E}_{\mathbf{k}\gamma}}), \\
  2u_{\mathbf{k}\gamma}v_{\mathbf{k}\gamma} =  \Delta_{\mathbf{k}\gamma}/\tilde{E}_{\mathbf{k}\gamma}.
 \end{align}

\subsection{Green's functions}
The following Kohn-Sham propagators enter the Feynman diagrams:
\begin{enumerate}
 \item The usual electron Green's function:
 \begin{align}
  G^e_{KS}(r\tau,r'\tau')=-\langle \hat{T} \hat{\psi}(r\tau)\hat{\psi}^\dagger(r'\tau') \rangle_{KS}
 \end{align}
 \item The hole Green's function:
 \begin{align}
  G^h_{KS}(r\tau,r'\tau')=-\langle \hat{T} \hat{\phi}^\dagger(r\tau)\hat{\phi}(r'\tau') \rangle_{KS}
 \end{align} 
  \item The anomalous Green's functions
 \begin{align}
  F^{he}_{KS}(r\tau,r'\tau')=-\langle \hat{T} \hat{\phi}(r\tau)\hat{\psi}(r'\tau') \rangle_{KS} \\
  F^{he}_{KS}\ ^\dagger(r\tau,r'\tau')=-\langle \hat{T} \hat{\psi}^\dagger(r'\tau')\hat{\phi}^\dagger(r\tau) \rangle_{KS} \\
  F^{eh}_{KS}(r\tau,r'\tau')=-\langle \hat{T} \hat{\psi}(r\tau)\hat{\phi}(r'\tau') \rangle_{KS} \\
  F^{eh}_{KS}\ ^\dagger(r\tau,r'\tau')=-\langle \hat{T} \hat{\phi}^\dagger(r'\tau')\hat{\psi}^\dagger(r\tau) \rangle_{KS}
  \end{align} 
\end{enumerate}
By inserting the Bogoliubov transformation in the expression for the propagators and also utilizing 
\begin{align}
&\langle \hat{T} \hat{\gamma} \hat{\gamma} \rangle = \langle \hat{T} \hat{\gamma}^\dagger \hat{\gamma}^\dagger \rangle = 0 \\
&\Gamma_{ab}(m,n,\tau-\tau') \equiv \langle \hat{T} \hat{\gamma}_{am}(\tau) \hat{\gamma}^\dagger_{bn}(\tau')\rangle = \delta_{mn}\delta_{ab} e^{-E_{am}(\tau-\tau')}\left[ \Theta(\tau-\tau')f_\beta(-E_{am}) - \Theta(\tau'-\tau)f_\beta(E_{am}\right]) \\
&\Gamma_{ab}^\dagger(m,n,\tau-\tau') \equiv \langle \hat{T} \hat{\gamma}^\dagger_{am}(\tau) \hat{\gamma}_{bn}(\tau')\rangle = \delta_{mn}\delta_{ab} e^{E_{am}(\tau-\tau')}\left[ \Theta(\tau-\tau')f_\beta(E_{am}) - \Theta(\tau'-\tau)f_\beta(-E_{am}\right]) \\
&\Gamma_{ab}(m,n,i\omega_n) = -\frac{\delta_{mn}\delta_{ab}}{i\omega_n - E_{am}} \\
&\Gamma^\dagger_{ab}(m,n,i\omega_n) = -\frac{\delta_{mn}\delta_{ab}}{i\omega_n + E_{am}}
\end{align}
(Note that $\Gamma_{ab}^\dagger(m,n,\tau-\tau') \neq (\Gamma_{ab}(m,n,\tau-\tau'))^\dagger$ but $\Gamma_{ab}^\dagger(m,n,\tau-\tau') = -(\Gamma_{ab}(m,n,\tau-\tau'))^\dagger$)
we arrive at \cite{ludersthesis}
\begin{align}
 G^e_{KS}(r\tau,r'\tau') =& -\langle \hat{T}\hat{\psi}(r\tau)\hat{\psi}^{\dagger}(r'\tau')\rangle_{KS} = -\sum_m \left[ u_m^e(r) u_m^{e*}(r') \Gamma_{11}(m,m,\tau-\tau') + v_m^{e*}(r)v_m^e(r')\Gamma^\dagger_{22}(m,m,\tau-\tau')\right] \\
 G^h_{KS}(r\tau,r'\tau') =& -\langle \hat{T}\hat{\phi}^\dagger(r\tau)\hat{\phi}(r'\tau')\rangle_{KS} = -\sum_m \left[ u_m^{h*}(r) u_m^h(r') \Gamma^\dagger_{22}(m,m,\tau-\tau') + v_m^{h}(r)v_m^{h*}(r')\Gamma_{11}(m,m,\tau-\tau')\right] \\ 
  F_{KS}^{he}(r\tau,r'\tau') =& -\langle \hat{T}\hat{\phi}(r\tau)\hat{\psi}(r'\tau')\rangle_{KS} = -\sum_m \left[  v_m^{h*}(r)u_m^e(r')\Gamma^\dagger_{11}(m,m,\tau-\tau') - u_m^h(r) v_m^{e*}(r') \Gamma_{22}(m,m,\tau-\tau')\right] \\
    F_{KS}^{eh}(r\tau,r'\tau') =& -\langle \hat{T}\hat{\psi}(r\tau)\hat{\phi}(r'\tau')\rangle_{KS} = -\sum_m \left[  v_m^{h*}(r')u_m^e(r)\Gamma_{11}(m,m,\tau-\tau') - u_m^h(r') v_m^{e*}(r) \Gamma_{22}^\dagger(m,m,\tau-\tau')\right]
\end{align}
with the corresponding Fourier transforms
\begin{align}
 G^e_{KS}(rr'i\omega_n) =& \sum_m \frac{u^e_m(r)u^{e*}_m(r')}{i\omega_n - \epsilon_{1m}} + \sum_m \frac{v^{e*}_m(r)v^{e}_m(r')}{i\omega_n + \epsilon_{2m}} \\
 G^h_{KS}(rr'i\omega_n) =& \sum_m \frac{u^{h*}_m(r)u^{h}_m(r')}{i\omega_n + \epsilon_{2m}} + \sum_m \frac{v^{h}_m(r)v^{h*}_m(r')}{i\omega_n - \epsilon_{1m}} \\
 F_{KS}^{he}(rr'i\omega_n) =& \sum_m \frac{v^{h*}_m(r)u^{e}_m(r')}{i\omega_n + \epsilon_{1m}} -\sum_m \frac{u^h_m(r)v^{e*}_m(r')}{i\omega_n - \epsilon_{2m}}\\
 F_{KS}^{eh}(rr'i\omega_n) =& \sum_m \frac{v^{h*}_m(r')u^{e}_m(r)}{i\omega_n - \epsilon_{1m}} -\sum_m \frac{u^h_m(r')v^{e*}_m(r)}{i\omega_n + \epsilon_{2m}} 
\end{align}

Within the decoupling approximation the propagators are given by
\begin{align}
 G^e_{KS}(rr'i\omega_n) =& \sum_m \frac{u^e_m(r)u^{e*}_m(r')}{i\omega_n - \epsilon_{1m}} + \sum_m \frac{v^{e*}_m(r)v^{e}_m(r')}{i\omega_n + \epsilon_{2m}} \approx 
 \sum_k  \psi_k(r)\psi_k^{*}(r') \overbrace{\left[ \frac{|u_k^e|^2}{i\omega_n - \epsilon_{1m}} + \frac{|v_k^e|^2}{i\omega_n + \epsilon_{2m}}\right]}^{G^e(k,\omega_n)} \\
 G^h_{KS}(rr'i\omega_n) =& \sum_m \frac{u^{h*}_m(r)u^{h}_m(r')}{i\omega_n + \epsilon_{2m}} + \sum_m \frac{v^{h}_m(r)v^{h*}_m(r')}{i\omega_n - \epsilon_{1m}} \approx
 \sum_k  \phi_{-k}(r)\phi_{-k}^{*}(r') \overbrace{\left[ \frac{|u_k^h|^2}{i\omega_n + \epsilon_{2m}} + \frac{|v_k^h|^2}{i\omega_n - \epsilon_{1m}}\right]}^{G^h(k,\omega_n)}\\
 F_{KS}^{eh}(rr'i\omega_n) =& \sum_m \frac{v^{h*}_m(r')u^{e}_m(r)}{i\omega_n - \epsilon_{1m}} -\sum_m \frac{u^h_m(r')v^{e*}_m(r)}{i\omega_n + \epsilon_{2m}} \approx
  \sum_k  \psi_k(r)\phi_{-k}^{*}(r') \overbrace{\left[ \frac{v_k^{h*}u_k^e}{i\omega_n - \epsilon_{1m}} - \frac{u_k^hv_k^{e*}}{i\omega_n + \epsilon_{2m}}\right]}^{F^{eh}(k,\omega_n)}
\end{align}

\subsection{Nambu-Gorkov Green's functions}
Before we proceed we have to define the Nambu-Gorkov propagators. First we define the Nambu-Gorkov field operators as
 \begin{gather}
 \bar{\Psi}_\sigma(\mathbf{r},\tau) = 
 \begin{bmatrix} \hat{\psi}_\sigma(\mathbf{r},\tau) \\ \hat{\phi}^\dagger_{-\sigma}(\mathbf{r},\tau)  \end{bmatrix}
 \end{gather}
 The Green's function is then the tensor product
 \begin{gather}
 \bar{G}_{\sigma \sigma'} (\mathbf{r}\tau,\mathbf{r}'\tau') = \langle \bar{\Psi}_\sigma(\mathbf{r},\tau) \otimes \bar{\Psi}_{\sigma'}(\mathbf{r}',\tau') \rangle =
 \begin{bmatrix} G^e_{\sigma \sigma'}(\mathbf{r}\tau,\mathbf{r}'\tau') &   F^{eh}_{\sigma,-\sigma'} (\mathbf{r}\tau,\mathbf{r}'\tau') \\ 
  -F^{he\dagger}_{-\sigma,\sigma'} (\mathbf{r}\tau,\mathbf{r}'\tau') &   G^h_{-\sigma', -\sigma}(\mathbf{r}\tau,\mathbf{r}'\tau') 
 \end{bmatrix}
 \end{gather} 

 The Kohn-Sham Nambu Gorkov Green's function obeys the equation of motion
 \begin{align}
  \hat{\mathcal{L}} \bar{G}^s_{\sigma \sigma'} (\mathbf{r}\tau,\mathbf{r}'\tau') = -\delta_{\sigma \sigma'} \delta(\mathbf{r}-\mathbf{r}')\delta(\tau-\tau')
 \end{align}
where
\begin{gather}
\hat{\mathcal{L}}=
 \begin{bmatrix} \frac{\partial}{\partial \tau} + H_e(\mathbf{r}) &  \hat{\Delta}_s(\mathbf{r}) \\ 
 \hat{\Delta}_s^*(\mathbf{r}) &  \frac{\partial}{\partial \tau} + H_h(\mathbf{r}) \end{bmatrix}
 \end{gather} 
 and 
 \begin{align}
   \hat{\Delta}_s(\mathbf{r})f(\mathbf{r}) = \int dr' \Delta_s(\mathbf{r},\mathbf{r}')f(\mathbf{r}')  
 \end{align}
 
\subsection{Self-energy}
We will divide the self-energy into the static part which comes from the exchange-correlation potentials and the rest 
\begin{align}
\bar{\Sigma}(\mathbf{r}\tau,\mathbf{r}'\tau') = \bar{\Sigma}_{xc}(\mathbf{r}\tau,\mathbf{r}'\tau') + \bar{U}(\mathbf{r},\mathbf{r}),
\label{SEdiv}
\end{align}
where
\begin{gather}
\bar{U}(\mathbf{r},\mathbf{r}') = -\delta_{\sigma \sigma'}
 \begin{bmatrix} \delta(\mathbf{r}-\mathbf{r}')v_{xc}^e(r) &  \hat{\Delta}_s(\mathbf{r},\mathbf{r}') \\ 
 \hat{\Delta}^*_s(\mathbf{r},\mathbf{r}') &  \delta(\mathbf{r}-\mathbf{r}')v_{xc}^h(r) \end{bmatrix}.
 \end{gather} 
The first order self-energy diagram in $W$ (corresponding to the $GW$-approximation) is given by
  \begin{gather}
\bar{\Sigma}^1_{xc}(\mathbf{r}\tau,\mathbf{r}'\tau') = W(\mathbf{r}\tau,\mathbf{r}'\tau')\times
 \begin{bmatrix} G^e_{\sigma \sigma'}(\mathbf{r}\tau,\mathbf{r}'\tau') &   F^{eh}_{\sigma,-\sigma'} (\mathbf{r}\tau,\mathbf{r}'\tau') \\ 
  -F^{he\dagger}_{-\sigma,\sigma'} (\mathbf{r}\tau,\mathbf{r}'\tau') &   G^h_{-\sigma', -\sigma}(\mathbf{r'}\tau',\mathbf{r}\tau) 
 \end{bmatrix} \end{gather}

\subsection{Sham-Schl\"{u}ter connection}
The The Sham-Schl\"{u}ter equations \cite{Godby86Accurate,Godby87Quasi,Godby87Trends} and the corresponding equations for 
superconducting systems developed by Marques et al.\cite{marquesthesis} make use of the equality between the Kohn-Sham and interacting densities:
\begin{align}
 n^{e/h}(\mathbf{r})=\pm \mathrm{lim}_{\eta\rightarrow0^+} \frac{1}{\beta}\sum_{\omega_n}\sum_{\sigma}e^{i \eta \omega_n}G^{e/h}_{\sigma \sigma}(\mathbf{rr};\pm \omega_n) = \pm \mathrm{lim}_{\eta\rightarrow0^+} \frac{1}{\beta}\sum_{\omega_n}\sum_{\sigma}e^{i \eta \omega_n}G^{e/h - s}_{\sigma \sigma}(\mathbf{rr};\pm \omega_n) \\
 \chi(\mathbf{r}',\mathbf{r})^*= -\mathrm{lim}_{\eta\rightarrow0^+} \frac{1}{\beta}\sum_{\omega_n}\sum_{\sigma}e^{i \eta \omega_n}F_{\sigma \sigma}^{eh}(\mathbf{rr}';-\omega_n) = -\mathrm{lim}_{\eta\rightarrow0^+} \frac{1}{\beta}\sum_{\omega_n}\sum_{\sigma}e^{i \eta \omega_n}F^{eh - s}_{\sigma \sigma}(\mathbf{rr}';-\omega_n) \\
 \chi(\mathbf{r}',\mathbf{r})= -\mathrm{lim}_{\eta\rightarrow0^+} \frac{1}{\beta}\sum_{\omega_n}\sum_{\sigma}e^{i \eta \omega_n}F_{\sigma \sigma}^{he}{}^\dagger(\mathbf{rr}';-\omega_n) = -\mathrm{lim}_{\eta\rightarrow0^+} \frac{1}{\beta}\sum_{\omega_n}\sum_{\sigma}e^{i \eta \omega_n}F^{he - s}{}^\dagger_{\sigma \sigma}(\mathbf{rr}';-\omega_n)
\end{align}

 Now using the Dyson equation and setting the interacting and Kohn-Sham densities equal we get the Sham-Schl\"{u}ter equations
 \begin{align}
  0=& \lim_{\eta \rightarrow 0^+} \frac{1}{\beta}\sum_{\omega_n} e^{i\eta \omega_n} \sum_{\sigma \sigma_1 \sigma_2} \int dr_1dr_2 
  [\bar{G}^s_{\sigma \sigma_1}(\mathbf{r}\mathbf{r}_1\omega_n) \bar{\Sigma}_{\sigma_1\sigma_2}(\mathbf{r}_1,\mathbf{r}_2\omega_n) \bar{G}_{\sigma_2 \sigma}(\mathbf{r}_2\mathbf{r}\omega_n)]_{11} \\
    0=& \lim_{\eta \rightarrow 0^+} \frac{1}{\beta}\sum_{\omega_n} e^{i\eta \omega_n} \sum_{\sigma \sigma_1 \sigma_2} \int dr_1dr_2 
  [\bar{G}^s_{\sigma \sigma_1}(\mathbf{r}\mathbf{r}_1;-\omega_n) \bar{\Sigma}_{\sigma_1\sigma_2}(\mathbf{r}_1,\mathbf{r}_2;-\omega_n) \bar{G}_{\sigma_2 \sigma}(\mathbf{r}_2\mathbf{r};-\omega_n)]_{22} \\
  0=& \lim_{\eta \rightarrow 0^+} \frac{1}{\beta}\sum_{\omega_n} e^{i\eta \omega_n} \sum_{\sigma_1 \sigma_2} \int dr_1dr_2 
  [\bar{G}^s_{\sigma \sigma_1}(\mathbf{r}\mathbf{r}_1;-\omega_n) \bar{\Sigma}_{\sigma_1\sigma_2}(\mathbf{r}_1\mathbf{r}_2;-\omega_n) 
  \bar{G}_{\sigma_2 \sigma'}(\mathbf{r}_2\mathbf{r}';-\omega_n)]_{12} \\
    0=& \lim_{\eta \rightarrow 0^+} \frac{1}{\beta}\sum_{\omega_n} e^{i\eta \omega_n} \sum_{\sigma_1 \sigma_2} \int dr_1dr_2 
  [\bar{G}^s_{\sigma \sigma_1}(\mathbf{r}\mathbf{r}_1;-\omega_n) \bar{\Sigma}_{\sigma_1\sigma_2}(\mathbf{r}_1\mathbf{r}_2;-\omega_n) 
  \bar{G}_{\sigma_2 \sigma'}(\mathbf{r}_2\mathbf{r}';-\omega_n)]_{21} 
  \end{align}
Using the division of the self-energy Eq. \ref{SEdiv} we get (henceforth dropping the index eh in $F$ with the implicit notation of $F=F^{eh}$ and $F^\dagger = F^{he\dagger}$)

 \begin{align}
  \frac{1}{\beta}&\sum_{\omega_n} \sum_{\sigma \sigma_1 \sigma_2} \int dr_1dr_2 
  [\bar{G}^s_{\sigma \sigma_1}(\mathbf{r}\mathbf{r}_1\omega_n) \bar{\Sigma}^{xc}_{\sigma_1\sigma_2}(\mathbf{r}_1,\mathbf{r}_2\omega_n) \bar{G}_{\sigma_2 \sigma}(\mathbf{r}_2\mathbf{r}\omega_n)]_{11} = \nonumber \\
  &\frac{2}{\beta}\sum_{\omega_n} \int dr_1 \left[ v_{xc}^e(r_1)G^e_s(r r_1;\omega_n)G^e(r_1r;\omega_n) + v_{xc}^h(r_1)F_s(rr_1;\omega_n)F^\dagger(r_1r;\omega_n)\right] + \nonumber \\
  &\frac{2}{\beta}\sum_{\omega_n} \int dr_1 dr_2 \left[ \Delta_{xc}^{*}(r_1r_2)F_s(r r_1;\omega_n)G^e(r_2r;\omega_n) + \Delta_{xc}(r_1,r_2)G^e_s(rr_1;\omega_n)F^\dagger(r_2r;\omega_n)\right] \\
  \frac{1}{\beta}&\sum_{\omega_n} \sum_{\sigma \sigma_1 \sigma_2} \int dr_1dr_2 
  [\bar{G}^s_{\sigma \sigma_1}(\mathbf{r}\mathbf{r}_1\omega_n) \bar{\Sigma}^{xc}_{\sigma_1\sigma_2}(\mathbf{r}_1,\mathbf{r}_2\omega_n) \bar{G}_{\sigma_2 \sigma}(\mathbf{r}_2\mathbf{r}\omega_n)]_{22} = \nonumber \\
  &\frac{2}{\beta}\sum_{\omega_n} \int dr_1 \left[ v_{xc}^h(r_1)G^h_s(r r_1;\omega_n)G^h(r_1r;\omega_n) + v_{xc}^e(r_1)F_s^\dagger(rr_1;\omega_n)F(r_1r;\omega_n)\right] + \nonumber \\
  &\frac{2}{\beta}\sum_{\omega_n} \int dr_1 dr_2 \left[ \Delta_{xc}(r_1r_2)F_s^\dagger(r r_1;\omega_n)G^h(r_2r;\omega_n) + \Delta_{xc}^*(r_1,r_2)G^h_s(rr_1;\omega_n)F(r_2r;\omega_n)\right] \\
  \frac{1}{\beta}&\sum_{\omega_n} \sum_{\sigma_1 \sigma_2} \int dr_1dr_2 
  [\bar{G}^s_{\sigma \sigma_1}(\mathbf{r}\mathbf{r}_1;\omega_n) \bar{\Sigma}^{xc}_{\sigma_1\sigma_2}(\mathbf{r}_1\mathbf{r}_2;\omega_n)  \bar{G}_{\sigma_2 \sigma'}(\mathbf{r}_2\mathbf{r}';\omega_n)]_{12} = \nonumber \\
  &\frac{1}{\beta}\sum_{\omega_n} \int dr_1 \left[ v_{xc}^e(r_1)G^e_s(r r_1;\omega_n)F(r_1r';\omega_n) + v_{xc}^h(r_1)F_s(rr_1;\omega_n)G^h(r_1r';\omega_n)\right] + \nonumber \\
  &\frac{1}{\beta}\sum_{\omega_n} \int dr_1 dr_2 \left[ \Delta_{xc}^{*}(r_1r_2)F_s(r r_1;\omega_n)F(r_2r';\omega_n) + \Delta_{xc}(r_1,r_2)G^e_s(rr_1;\omega_n)G^h(r_2r';\omega_n)\right] \\
  \frac{1}{\beta}&\sum_{\omega_n} \sum_{\sigma_1 \sigma_2} \int dr_1dr_2 
  [\bar{G}^s_{\sigma \sigma_1}(\mathbf{r}\mathbf{r}_1;\omega_n) \bar{\Sigma}^{xc}_{\sigma_1\sigma_2}(\mathbf{r}_1\mathbf{r}_2;\omega_n)  \bar{G}_{\sigma_2 \sigma'}(\mathbf{r}_2\mathbf{r}';\omega_n)]_{21} = \nonumber \\
  &\frac{1}{\beta}\sum_{\omega_n} \int dr_1 \left[ v_{xc}^h(r_1)G^h_s(r r_1;\omega_n)F^\dagger(r_1r';\omega_n) + v_{xc}^e(r_1)F_s^\dagger(rr_1;\omega_n)G^e(r_1r';\omega_n)\right] + \nonumber \\
  &\frac{1}{\beta}\sum_{\omega_n} \int dr_1 dr_2 \left[ \Delta_{xc}(r_1r_2)F_s^\dagger(r r_1;\omega_n)F^\dagger(r_2r';\omega_n) + \Delta^*_{xc}(r_1,r_2)G^h_s(rr_1;\omega_n)G^e(r_2r';\omega_n)\right]
  \end{align}
where we also have used the fact that $\sum_{\omega_n} f(-\omega_n)= \sum_{\omega_n} f(\omega_n)$.
Now we write out the first line in each equation explicitly assuming that the self-energy does not couple different spin-channels in Nambu-Gorkov space

 \begin{align}
 \frac{1}{\beta}&\sum_{\omega_n} \int dr_1 dr_2 \left[ \Sigma^{xc}_{11}(r_1,r_2,\omega_n)G^e_s(r r_1;\omega_n)G^e(r_2r;\omega_n) + \Sigma^{xc}_{22}(r_1,r_2)F_s(rr_1;\omega_n)F^\dagger(r_2r;\omega_n)\right] + \nonumber \\
  &\frac{1}{\beta}\sum_{\omega_n} \int dr_1 dr_2 \left[ \Sigma^{xc}_{21}(r_1r_2 \omega_n)F_s(r r_1;\omega_n)G^e(r_2r;\omega_n) + \Sigma^{xc}_{12}(r_1,r_2,\omega_n)G^e_s(rr_1;\omega_n)F^\dagger(r_2r;\omega_n)\right] = \nonumber \\
  &\frac{1}{\beta}\sum_{\omega_n} \int dr_1 \left[ v_{xc}^e(r_1)G^e_s(r r_1;\omega_n)G^e(r_1r;\omega_n) + v_{xc}^h(r_1)F_s(rr_1;\omega_n)F^\dagger(r_1r;\omega_n)\right] + \nonumber \\
  &\frac{1}{\beta}\sum_{\omega_n} \int dr_1 dr_2 \left[ \Delta_{xc}^{*}(r_1r_2)F_s(r r_1;\omega_n)G^e(r_2r;\omega_n) + \Delta_{xc}(r_1,r_2)G^e_s(rr_1;\omega_n)F^\dagger(r_2r;\omega_n)\right] \\
  \frac{1}{\beta}&\sum_{\omega_n} \int dr_1 dr_2 \left[ \Sigma^{xc}_{22}(r_1,r_2,\omega_n)G^h_s(r r_1;\omega_n)G^h(r_2r;\omega_n) + \Sigma^{xc}_{11}(r_1,r_2,\omega_n)F_s^\dagger(rr_1;\omega_n)F(r_2r;\omega_n)\right] + \nonumber \\
  &\frac{1}{\beta}\sum_{\omega_n} \int dr_1 dr_2 \left[ \Sigma^{xc}_{12}(r_1r_2,\omega_n)F_s^\dagger(r r_1;\omega_n)G^h(r_2r;\omega_n) + \Sigma^{xc}_{21}(r_1,r_2,\omega_n)G^h_s(rr_1;\omega_n)F(r_2r;\omega_n)\right] \nonumber = \\ 
  &\frac{1}{\beta}\sum_{\omega_n} \int dr_1 \left[ v_{xc}^h(r_1)G^h_s(r r_1;\omega_n)G^h(r_1r;\omega_n) + v_{xc}^e(r_1)F_s^\dagger(rr_1;\omega_n)F(r_1r;\omega_n)\right] + \nonumber \\
  &\frac{1}{\beta}\sum_{\omega_n} \int dr_1 dr_2 \left[ \Delta_{xc}(r_1r_2)F_s^\dagger(r r_1;\omega_n)G^h(r_2r;\omega_n) + \Delta_{xc}^*(r_1,r_2)G^h_s(rr_1;\omega_n)F(r_2r;\omega_n)\right] \\
  \frac{1}{\beta}&\sum_{\omega_n} \int dr_1 dr_2 \left[ \Sigma^{xc}_{11}(r_1,r_2,\omega_n)G^e_s(r r_1;\omega_n)F(r_2r';\omega_n) + \Sigma^{xc}_{22}(r_1,r_2,\omega_n)F_s(rr_1;\omega_n)G^h(r_2r';\omega_n)\right] + \nonumber \\
  &\frac{1}{\beta}\sum_{\omega_n} \int dr_1 dr_2 \left[ \Sigma^{xc}_{21}(r_1,r_2,\omega_n)F_s(r r_1;\omega_n)F(r_2r';\omega_n) + \Sigma^{xc}_{12}(r_1,r_2,\omega_n)G^e_s(rr_1;\omega_n)G^h(r_2r';\omega_n)\right] = \nonumber \\
  &\frac{1}{\beta}\sum_{\omega_n} \int dr_1 \left[ v_{xc}^e(r_1)G^e_s(r r_1;\omega_n)F(r_1r';\omega_n) + v_{xc}^h(r_1)F_s(rr_1;\omega_n)G^h(r_1r';\omega_n)\right] + \nonumber \\
  &\frac{1}{\beta}\sum_{\omega_n} \int dr_1 dr_2 \left[ \Delta_{xc}^{*}(r_1r_2)F_s(r r_1;\omega_n)F(r_2r';\omega_n) - \Delta_{xc}(r_1,r_2)G^e_s(rr_1;\omega_n)G^h(r_2r';\omega_n)\right] \\
  \frac{1}{\beta}&\sum_{\omega_n} \int dr_1 dr_2 \left[ \Sigma^{xc}_{22}(r_1,r_2,\omega_n)G^h_s(r r_1;\omega_n)F^\dagger(r_2r';\omega_n) + \Sigma^{xc}_{11}(r_1,r_2,\omega_n)F_s^\dagger(rr_1;\omega_n)G^e(r_2r';\omega_n)\right] + \nonumber \\
  &\frac{1}{\beta}\sum_{\omega_n} \int dr_1 dr_2 \left[ \Sigma^{xc}_{12}(r_1,r_2,\omega_n)F_s^\dagger(r r_1;\omega_n)F^\dagger(r_2r';\omega_n) + \Sigma^{xc}_{21}(r_1,r_2,\omega_n)G^h_s(rr_1;\omega_n)G^e(r_2r';\omega_n)\right] = \nonumber \\
  &\frac{1}{\beta}\sum_{\omega_n} \int dr_1 \left[ v_{xc}^h(r_1)G^h_s(r r_1;\omega_n)F^\dagger(r_1r';\omega_n) + v_{xc}^e(r_1)F_s^\dagger(rr_1;\omega_n)G^e(r_1r';\omega_n)\right] + \nonumber \\
  &\frac{1}{\beta}\sum_{\omega_n} \int dr_1 dr_2 \left[ \Delta_{xc}(r_1r_2)F_s^\dagger(r r_1;\omega_n)F^\dagger(r_2r';\omega_n) + \Delta^*_{xc}(r_1,r_2)G^h_s(rr_1;\omega_n)G^e(r_2r';\omega_n)\right]
  \end{align}

To simplify the equations we make use of the decoupling approximation and take the matrix-elements of the equations. We will assume time-reversal symmetry, which implies that the LDA eigenstates are real. Furthermore, to simplify the notation we will ignore the band-index in the derivation below. However, it should be noted that the band-index enters in the same way as the $k$-index, so that the final gap equation is valid also for the multiorbital models.
We can then write the Green's function as

\begin{align}
 G^e_{KS}(rr'i\omega_n) =& \sum_m \frac{u^e_m(r)u^{e*}_m(r')}{i\omega_n - \epsilon_{1m}} + \sum_m \frac{v^{e*}_m(r)v^{e}_m(r')}{i\omega_n + \epsilon_{2m}} \approx 
 \sum_\mathbf{k}  \psi_\mathbf{k}(r)\psi_\mathbf{k}^{*}(r') \overbrace{\left[ \frac{|u_\mathbf{k}|^2}{i\omega_n - \epsilon_{1\mathbf{k}}} + \frac{|v_\mathbf{k}|^2}{i\omega_n + \epsilon_{2\mathbf{k}}}\right]}^{G^e(\mathbf{k},\omega_n)} \\
 G^h_{KS}(rr'i\omega_n) =& \sum_m \frac{u^{h*}_m(r)u^{h}_m(r')}{i\omega_n + \epsilon_{2m}} + \sum_m \frac{v^{h}_m(r)v^{h*}_m(r')}{i\omega_n - \epsilon_{1m}} \approx
 \sum_\mathbf{k}  \phi_{-\mathbf{k}}(r)\phi_{-\mathbf{k}}^{*}(r') \overbrace{\left[ \frac{|u_\mathbf{k}|^2}{i\omega_n + \epsilon_{2\mathbf{k}}} + \frac{|v_\mathbf{k}|^2}{i\omega_n - \epsilon_{1\mathbf{k}}}\right]}^{G^h(\mathbf{k},\omega_n)}\\
 F_{KS}^{eh}(rr'i\omega_n) =& \sum_m \frac{v^{h*}_m(r')u^{e}_m(r)}{i\omega_n - \epsilon_{1m}} -\sum_m \frac{u^h_m(r')v^{e*}_m(r)}{i\omega_n + \epsilon_{2m}} \approx
  \sum_\mathbf{k}  \psi_\mathbf{k}(r)\phi_{-\mathbf{k}}^{*}(r') \overbrace{\left[ \frac{v_\mathbf{k}^{*}u_\mathbf{k}}{i\omega_n - \epsilon_{1\mathbf{k}}} - \frac{u_\mathbf{k}v_k^{*}}{i\omega_n + \epsilon_{2\mathbf{k}}}\right]}^{F^{eh}(\mathbf{k},\omega_n)}
\end{align}

Furthermore we define
\begin{align}
\Sigma^{12}(\mathbf{k},\mathbf{k}',\omega_n) = \int dr dr' \psi^*_\mathbf{k}(r)\Sigma^{12}(rr'\omega_n)\phi_{-\mathbf{k}'}(r') \\
\Sigma^{21}(\mathbf{k},\mathbf{k}',\omega_n) = \int dr dr' \phi^*_{-\mathbf{k}}(r)\Sigma^{21}(rr'\omega_n)\psi_{\mathbf{k}'}(r') \\
\Sigma^{22}(\mathbf{k},\mathbf{k}',\omega_n) = \int dr dr' \phi^*_{-\mathbf{k}}(r)\Sigma^{22}(rr'\omega_n)\phi_{-\mathbf{k}'}(r') \\
\Sigma^{11}(\mathbf{k},\mathbf{k}',\omega_n) = \int dr dr' \psi^*_{\mathbf{k}}(r)\Sigma^{11}(rr'\omega_n)\psi_{\mathbf{k}'}(r')
\end{align}
and so on for the different electron-electron, electron-hole, hole-hole and hole-electron components. Correspondingly for $v_{xc}$ we define
\begin{align}
v_{xc}^{e}(\mathbf{k},\mathbf{k}') = \int dr \psi^*_\mathbf{k}(r)v_{xc}^{e}(r)\psi_{\mathbf{k}'}(r) \\
v_{xc}^{h}(\mathbf{k},\mathbf{k}') = \int dr \phi^*_{-\mathbf{k}}(r)v_{xc}^{h}(r)\phi_{-\mathbf{k}'}(r) \\
\Delta(\mathbf{k}) =  \int dr dr' \psi^{*}_\mathbf{k}(r)\Delta(r,r')\phi_{-\mathbf{k}}(r')
\end{align}
Any matrix element with $v_{xc}^e$ including hole wavefunctions is set to zero and vice versa. 

Now we insert these expressions into the equations above. In the diagonal components, which are derived from the equations of the normal density we integrate over $\mathbf{r}$ and use the orthonormality of the wave-functions
to get the equations for the single components. Furthermore in the off-diagonal components, which are derived from the anomalous density we multiply by $\psi_\mathbf{k}(r)$ and $\phi^*_{-\mathbf{k}}(r')$ and integrate over $\mathbf{r}$ and 
$\mathbf{r}'$ in the third equation and the other way around in the last equation to get the equations for the single components.

 \begin{align}
 \sum_{\omega_n,\mathbf{k}} &[ \Sigma^{xc}_{11}(\mathbf{k},\mathbf{k},\omega_n)G^e_s(\mathbf{k};\omega_n)G^e(\mathbf{k};\omega_n) + \Sigma^{xc}_{22}(\mathbf{k},\mathbf{k},\omega_n)F_s(\mathbf{k};\omega_n)F^\dagger(\mathbf{k};\omega_n) + \nonumber \\
 &\Sigma^{xc}_{21}(\mathbf{k},\mathbf{k} \omega_n)F_s(\mathbf{k};\omega_n)G^e(\mathbf{k};\omega_n) + \Sigma^{xc}_{12}(\mathbf{k},\mathbf{k} \omega_n)G_s^e(\mathbf{k};\omega_n)F^\dagger(\mathbf{k};\omega_n) ] = \nonumber \\
  \sum_{\omega_n\mathbf{k}}  &[v_{xc}^e(\mathbf{k},\mathbf{k})G^e_s(\mathbf{k};\omega_n)G^e(\mathbf{k};\omega_n) +v_{xc}^h(\mathbf{k},\mathbf{k})F_s(\mathbf{k};\omega_n)F^\dagger(\mathbf{k};\omega_n) + \nonumber \\
   &\Delta_{xc}^{*}(\mathbf{k})F_s(\mathbf{k};\omega_n)G^e(\mathbf{k};\omega_n) + \Delta_{xc}(\mathbf{k})G^e_s(\mathbf{k};\omega_n)F^\dagger(\mathbf{k};\omega_n)
   \label{A}
 \end{align}
 \vspace{1cm}
 \begin{align}
 \sum_{\omega_n\mathbf{k}} &[ \Sigma^{xc}_{22}(\mathbf{k}\mathbf{k},\omega_n)G^h_s(\mathbf{k};\omega_n)G^h(\mathbf{k};\omega_n) +  \Sigma^{xc}_{11}(\mathbf{k}\mathbf{k},\omega_n)F^\dagger_s(\mathbf{k};\omega_n)F(\mathbf{k};\omega_n)  + \nonumber \\
   &\Sigma^{xc}_{21}(\mathbf{k},\mathbf{k},\omega_n)G^h_s(\mathbf{k};\omega_n)F(\mathbf{k};\omega_n) + \Sigma^{xc}_{12}(\mathbf{k},\mathbf{k},\omega_n)F_s^\dagger(\mathbf{k};\omega_n)G^h(\mathbf{k};\omega_n) ] \nonumber  = \\ 
  \sum_{\omega_n,\mathbf{k}}  &[v_{xc}^h(\mathbf{k},\mathbf{k})G^h_s(\mathbf{k};\omega_n)G^h(\mathbf{k};\omega_n) + v_{xc}^e(\mathbf{k},\mathbf{k})F_s^\dagger(\mathbf{k};\omega_n)F(\mathbf{k};\omega_n)  + \nonumber \\
  &\Delta_{xc}(\mathbf{k})F^\dagger_s(\mathbf{k};\omega_n)G^h(\mathbf{k};\omega_n) + \Delta_{xc}^*(\mathbf{k})G^h_s(\mathbf{k};\omega_n)F(\mathbf{k};\omega_n) ]   
   \end{align}
 \vspace{1cm}
 \begin{align}
 \sum_{\omega_n} &[ \Sigma^{xc}_{11}(\mathbf{k},\mathbf{k}',\omega_n)G^e_s(\mathbf{k};\omega_n)F(\mathbf{k}';\omega_n) + \Sigma^{xc}_{22}(\mathbf{k},\mathbf{k}',\omega_n)F_s(\mathbf{k};\omega_n)G^h(\mathbf{k}';\omega_n) + \nonumber \\
   &\Sigma^{xc}_{21}(\mathbf{k},\mathbf{k}',\omega_n)F_s(\mathbf{k};\omega_n)F(\mathbf{k}';\omega_n) + \Sigma^{xc}_{12}(\mathbf{k},\mathbf{k}',\omega_n)G^e_s(\mathbf{k};\omega_n)G^h(\mathbf{k}';\omega_n) ] = \nonumber \\
 \sum_{\omega_n} &[ v_{xc}^e(\mathbf{k},\mathbf{k}')G^e_s(\mathbf{k};\omega_n)F(\mathbf{k}';\omega_n) + v_{xc}^h(\mathbf{k},\mathbf{k}')F_s(\mathbf{k};\omega_n)G^h(\mathbf{k}';\omega_n) + \nonumber \\
  &\delta_{\mathbf{k}\mathbf{k}'}\Delta_{xc}^{*}(\mathbf{k})F_s(\mathbf{k};\omega_n)F(\mathbf{k}';\omega_n) + \delta_{\mathbf{k}\mathbf{k}'}\Delta_{xc}(\mathbf{k})G^e_s(\mathbf{k};\omega_n)G^h(\mathbf{k}';\omega_n) ]
   \end{align}
 \vspace{1cm}
 \begin{align}
  \sum_{\omega_n} &[\Sigma^{xc}_{22}(\mathbf{k},\mathbf{k}',\omega_n)G^h_s(\mathbf{k};\omega_n)F^\dagger(\mathbf{k}';\omega_n) + \Sigma^{xc}_{11}(\mathbf{k},\mathbf{k}',\omega_n)F_s^\dagger(\mathbf{k};\omega_n)G^e(\mathbf{k}';\omega_n) + \nonumber \\
  &\Sigma^{xc}_{12}(\mathbf{k},\mathbf{k}',\omega_n)F_s^\dagger(\mathbf{k};\omega_n)F^\dagger(\mathbf{k}';\omega_n) + \Sigma^{xc}_{21}(\mathbf{k},\mathbf{k}',\omega_n)G^h_s(\mathbf{k};\omega_n)G^e(\mathbf{k}';\omega_n) ] = \nonumber \\
  \sum_{\omega_n} &[ v_{xc}^h(\mathbf{k},\mathbf{k}')G^h_s(\mathbf{k};\omega_n)F^\dagger(\mathbf{k}';\omega_n) + v_{xc}^e(\mathbf{k},\mathbf{k}')F_s^\dagger(\mathbf{k};\omega_n)G^e(\mathbf{k};\omega_n) \nonumber \\
  &\delta_{\mathbf{k}\mathbf{k}'}\Delta_{xc}(\mathbf{k})F_s^\dagger(\mathbf{k};\omega_n)F^\dagger(\mathbf{k}';\omega_n) + \delta_{\mathbf{k}\mathbf{k}'}\Delta^*_{xc}(\mathbf{k})G^h_s(\mathbf{k};\omega_n)G^e(\mathbf{k}';\omega_n) ]
  \label{C}
    \end{align}

   Now we set $\mathbf{k}'=\mathbf{k}$ in the equations above. Furthermore we ignore the $\mathbf{k}$-dependence of $v_{xc}$, i.e. $v_{xc}(\mathbf{k},\mathbf{k}')=v_{xc}$. This approximation is also done in the Sham-Shl\"uter derivation of scDFT.
 \begin{align}
 \sum_{\omega_n,\mathbf{k}} &[ \Sigma^{xc}_{11}(\mathbf{k},\mathbf{k},\omega_n)G^e_s(\mathbf{k};\omega_n)G^e(\mathbf{k};\omega_n) + \Sigma^{xc}_{22}(\mathbf{k},\mathbf{k},\omega_n)F_s(\mathbf{k};\omega_n)F^\dagger(\mathbf{k};\omega_n) + \nonumber \\
 &\Sigma^{xc}_{21}(\mathbf{k},\mathbf{k} \omega_n)F_s(\mathbf{k};\omega_n)G^e(\mathbf{k};\omega_n) + \Sigma^{xc}_{12}(\mathbf{k},\mathbf{k} \omega_n)G_s^e(\mathbf{k};\omega_n)F^\dagger(\mathbf{k};\omega_n) ] = \nonumber \\
  \sum_{\omega_n\mathbf{k}}  &[v_{xc}^e G^e_s(\mathbf{k};\omega_n)G^e(\mathbf{k};\omega_n) +v_{xc}^h F_s(\mathbf{k};\omega_n)F^\dagger(\mathbf{k};\omega_n) + \nonumber \\
   &\Delta_{xc}^{*}(\mathbf{k})F_s(\mathbf{k};\omega_n)G^e(\mathbf{k};\omega_n) + \Delta_{xc}(\mathbf{k})G^e_s(\mathbf{k};\omega_n)F^\dagger(\mathbf{k};\omega_n)]
   \label{A2}
 \end{align}
 \vspace{1cm}
 \begin{align}
 \sum_{\omega_n\mathbf{k}} &[ \Sigma^{xc}_{22}(\mathbf{k}\mathbf{k},\omega_n)G^h_s(\mathbf{k};\omega_n)G^h(\mathbf{k};\omega_n) +  \Sigma^{xc}_{11}(\mathbf{k}\mathbf{k},\omega_n)F^\dagger_s(\mathbf{k};\omega_n)F(\mathbf{k};\omega_n)  + \nonumber \\
   &\Sigma^{xc}_{21}(\mathbf{k},\mathbf{k},\omega_n)G^h_s(\mathbf{k};\omega_n)F(\mathbf{k};\omega_n) + \Sigma^{xc}_{12}(\mathbf{k},\mathbf{k},\omega_n)F_s^\dagger(\mathbf{k};\omega_n)G^h(\mathbf{k};\omega_n) ] \nonumber  = \\ 
  \sum_{\omega_n,\mathbf{k}}  &[v_{xc}^h G^h_s(\mathbf{k};\omega_n)G^h(\mathbf{k};\omega_n) + v_{xc}^e F_s^\dagger(\mathbf{k};\omega_n)F(\mathbf{k};\omega_n)  + \nonumber \\
  &\Delta_{xc}(\mathbf{k})F^\dagger_s(\mathbf{k};\omega_n)G^h(\mathbf{k};\omega_n) + \Delta_{xc}^*(\mathbf{k})G^h_s(\mathbf{k};\omega_n)F(\mathbf{k};\omega_n) ]   
   \end{align}
 \vspace{1cm}
 \begin{align}
 \sum_{\omega_n} &[ \Sigma^{xc}_{11}(\mathbf{k},\mathbf{k},\omega_n)G^e_s(\mathbf{k};\omega_n)F(\mathbf{k};\omega_n) + \Sigma^{xc}_{22}(\mathbf{k},\mathbf{k},\omega_n)F_s(\mathbf{k};\omega_n)G^h(\mathbf{k};\omega_n) + \nonumber \\
   &\Sigma^{xc}_{21}(\mathbf{k},\mathbf{k},\omega_n)F_s(\mathbf{k};\omega_n)F(\mathbf{k};\omega_n) + \Sigma^{xc}_{12}(\mathbf{k},\mathbf{k},\omega_n)G^e_s(\mathbf{k};\omega_n)G^h(\mathbf{k};\omega_n) ] = \nonumber \\
 \sum_{\omega_n} &[ v_{xc}^e G^e_s(\mathbf{k};\omega_n)F(\mathbf{k};\omega_n) + v_{xc}^h F_s(\mathbf{k};\omega_n)G^h(\mathbf{k};\omega_n) + \nonumber \\
  &\Delta_{xc}^{*}(\mathbf{k})F_s(\mathbf{k};\omega_n)F(\mathbf{k};\omega_n) + \Delta_{xc}(\mathbf{k})G^e_s(\mathbf{k};\omega_n)G^h(\mathbf{k};\omega_n) ]
   \end{align}
 \vspace{1cm}
 \begin{align}
  \sum_{\omega_n} &[\Sigma^{xc}_{22}(\mathbf{k},\mathbf{k},\omega_n)G^h_s(\mathbf{k};\omega_n)F^\dagger(\mathbf{k};\omega_n) + \Sigma^{xc}_{11}(\mathbf{k},\mathbf{k},\omega_n)F_s^\dagger(\mathbf{k};\omega_n)G^e(\mathbf{k};\omega_n) + \nonumber \\
  &\Sigma^{xc}_{12}(\mathbf{k},\mathbf{k},\omega_n)F_s^\dagger(\mathbf{k};\omega_n)F^\dagger(\mathbf{k};\omega_n) + \Sigma^{xc}_{21}(\mathbf{k},\mathbf{k},\omega_n)G^h_s(\mathbf{k};\omega_n)G^e(\mathbf{k};\omega_n) ] = \nonumber \\
  \sum_{\omega_n} &[ v_{xc}^h G^h_s(\mathbf{k};\omega_n)F^\dagger(\mathbf{k};\omega_n) + v_{xc}^e F_s^\dagger(\mathbf{k};\omega_n)G^e(\mathbf{k};\omega_n) \nonumber \\
  &\Delta_{xc}(\mathbf{k})F_s^\dagger(\mathbf{k};\omega_n)F^\dagger(\mathbf{k};\omega_n) + \Delta^*_{xc}(\mathbf{k})G^h_s(\mathbf{k};\omega_n)G^e(\mathbf{k};\omega_n) ]
  \label{C2}
    \end{align}

We now make a linear approximation in $\Delta$.
    \begin{align}
 \sum_{\omega_n,\mathbf{k}}  \Sigma^{xc}_{11}(\mathbf{k},\mathbf{k},\omega_n)G^e_s(\mathbf{k};\omega_n)G^e(\mathbf{k};\omega_n) = 
  \sum_{\omega_n\mathbf{k}}  v_{xc}^e G^e_s(\mathbf{k};\omega_n)G^e(\mathbf{k};\omega_n) 
 \end{align}
 \vspace{1cm}
 \begin{align}
 \sum_{\omega_n\mathbf{k}}  \Sigma^{xc}_{22}(\mathbf{k}\mathbf{k},\omega_n)G^h_s(\mathbf{k};\omega_n)G^h(\mathbf{k};\omega_n) =
\sum_{\omega_n,\mathbf{k}}  v_{xc}^h G^h_s(\mathbf{k};\omega_n)G^h(\mathbf{k};\omega_n)    
   \end{align}
 \vspace{1cm}
 \begin{align}
 \sum_{\omega_n} &[ \Sigma^{xc}_{11}(\mathbf{k},\mathbf{k},\omega_n)G^e_s(\mathbf{k};\omega_n)F(\mathbf{k};\omega_n) + \Sigma^{xc}_{22}(\mathbf{k},\mathbf{k},\omega_n)F_s(\mathbf{k};\omega_n)G^h(\mathbf{k};\omega_n) + \nonumber \\
   &\Sigma^{xc}_{12}(\mathbf{k},\mathbf{k},\omega_n)G^e_s(\mathbf{k};\omega_n)G^h(\mathbf{k};\omega_n) ] = \nonumber \\
 \sum_{\omega_n} &[ v_{xc}^e G^e_s(\mathbf{k};\omega_n)F(\mathbf{k};\omega_n) + v_{xc}^h F_s(\mathbf{k};\omega_n)G^h(\mathbf{k};\omega_n) + \nonumber \\
  &\Delta_{xc}(\mathbf{k})G^e_s(\mathbf{k};\omega_n)G^h(\mathbf{k};\omega_n) ]
   \end{align}
 \vspace{1cm}
 \begin{align}
  \sum_{\omega_n} &[\Sigma^{xc}_{22}(\mathbf{k},\mathbf{k},\omega_n)G^h_s(\mathbf{k};\omega_n)F^\dagger(\mathbf{k};\omega_n) + \Sigma^{xc}_{11}(\mathbf{k},\mathbf{k},\omega_n)F_s^\dagger(\mathbf{k};\omega_n)G^e(\mathbf{k};\omega_n) + \nonumber \\
  &\Sigma^{xc}_{21}(\mathbf{k},\mathbf{k},\omega_n)G^h_s(\mathbf{k};\omega_n)G^e(\mathbf{k};\omega_n) ] = \nonumber \\
  \sum_{\omega_n} &[ v_{xc}^h G^h_s(\mathbf{k};\omega_n)F^\dagger(\mathbf{k};\omega_n) + v_{xc}^e F_s^\dagger(\mathbf{k};\omega_n)G^e(\mathbf{k};\omega_n) \nonumber \\
  &\Delta^*_{xc}(\mathbf{k})G^h_s(\mathbf{k};\omega_n)G^e(\mathbf{k};\omega_n) ]
  \label{C}
    \end{align}

 Eliminating  $v_{xc}^e$ and $v_{xc}^h$ yields a linear equation for $\Delta$
  \begin{align}
 &\Delta^*_{xc}(\mathbf{k})\sum_{\omega_n} [G^h_s(\mathbf{k};\omega_n)G^e(\mathbf{k};\omega_n) ] = \nonumber \\  
&\sum_{\omega_n} [\Sigma^{xc}_{22}(\mathbf{k},\mathbf{k},\omega_n)G^h_s(\mathbf{k};\omega_n)F^\dagger(\mathbf{k};\omega_n) + \Sigma^{xc}_{11}(\mathbf{k},\mathbf{k},\omega_n)F_s^\dagger(\mathbf{k};\omega_n)G^e(\mathbf{k};\omega_n) + \nonumber \\
  &\Sigma^{xc}_{21}(\mathbf{k},\mathbf{k},\omega_n)G^h_s(\mathbf{k};\omega_n)G^e(\mathbf{k};\omega_n) ]  \nonumber \\
&-\sum_{\omega_n\mathbf{k}}  \Sigma^{xc}_{22}(\mathbf{k}\mathbf{k},\omega_n)G^h_s(\mathbf{k};\omega_n)G^h(\mathbf{k};\omega_n)
  \frac{\sum_{\omega_n} G^h_s(\mathbf{k};\omega_n)F^\dagger(\mathbf{k};\omega_n)}{\sum_{\omega_n\mathbf{k}}G^h_s(\mathbf{k};\omega_n)G^h(\mathbf{k};\omega_n) } \nonumber \\
  &-\sum_{\omega_n,\mathbf{k}}  \Sigma^{xc}_{11}(\mathbf{k},\mathbf{k},\omega_n)G^e_s(\mathbf{k};\omega_n)G^e(\mathbf{k};\omega_n) 
   \frac{\sum_{\omega_n}  F_s^\dagger(\mathbf{k};\omega_n)G^e(\mathbf{k};\omega_n)}{\sum_{\omega_n\mathbf{k}}   G^e_s(\mathbf{k};\omega_n)G^e(\mathbf{k};\omega_n)}
 \label{Eq:gapfull}
 \end{align}

 This reduces to the mean-field result if we set $\Sigma_{22}=\Sigma_{11}=0$ and assume that $\Sigma_{21}$ is 
 a static exchange term,

 \begin{align}
 \Delta_{\mathbf{k}\gamma}=\Sigma^{xc}_{12}(\gamma,\mathbf{k})=\Sigma^{xc}_{21}(\gamma,\mathbf{k}) &=  -\sum_{\mathbf{k}',\gamma'}\mathcal{F}_{\mathbf{k}\mathbf{k}'}^{\gamma\gamma'}\frac{V_{\mathbf{k}-\mathbf{k}'}\Delta_{\gamma'\mathbf{k}'}}{2E_{\gamma'\mathbf{k}'}}(1-f_\beta(\epsilon_{\gamma'\mathbf{k}'+}) - f_\beta(\epsilon_{\gamma'\mathbf{k}'-})).
 \end{align}

where $\mathcal{F}_{\mathbf{k}\mathbf{k}'}^{\gamma\gamma'}$ is the geometrical formfactor (defined below).  

\subsection{Further approximations}
For practical calculations we will assume that the bare dispersion already includes the effect of the normal self-energies and therefore set $\Sigma_{11}=\Sigma_{22}=0$ in Eq. \ref{Eq:gapfull}. Furthermore we will approximate the interacting propagators in the gap equation by their Kohn-Sham counterparts and evaluate the anomalous self-energy within the $GW$-approximation.

\subsection{Formfactor}
\label{sec:ff}
The geometrical form-factor $\mathcal{F}$ is related to the overlap of the single particle states $|f_{\mathbf{k}\gamma}\rangle$ as $\mathcal{F}_{\mathbf{k}\mathbf{k}'}^{\gamma\gamma'} = \langle f_{\mathbf{k},\gamma} | f_{\mathbf{k}',\gamma'} \rangle \langle f_{\mathbf{k}',\gamma'} | f_{\mathbf{k},\gamma} \rangle$ \cite{lozovik2010ultra}. We use the bilayer bandstructure defined as
\cite{Partoens2006,mccann2013,Conti2019}
\begin{align}
\varepsilon_{\mathbf{k}\gamma} = \frac{\gamma}{2}\sqrt{(t_1-\Gamma_k)^2 + \Omega_k}  - \mu  
\end{align}
where
\begin{align*}
    &\Gamma_k=\sqrt{t_1^2 + (2\hbar v k)^2 + (2E_g\hbar v k)^2/t_1^2}, \\
    &\Omega_k= E_g^2 (1-(2\hbar vk)^2/t_1^2), \\
    &v=\sqrt{3} a t_0 / 2\hbar
\end{align*}
where $a$ is the intercell distance, $t_0$ is the intralayer  hopping and $t_1$ is the interlayer  hopping. The bilayer bandstructure correspond to the two bands closest to the Fermi energy of the 4x4 Hamiltonian in Eq 30 of Ref. \onlinecite{mccann2013}. (See Eq. 43, 31, 30 with $\gamma_1=t_1$, $U=E_g$, $\alpha=1$, $v_3=v_4=0$, $\epsilon_{A1}=\epsilon_{B1}=-U/2$, $\epsilon_{A2}=\epsilon_{B2}=U/2$).

\begin{gather}
 H=\begin{bmatrix}-\frac{U}{2} & v\pi^+ & 0 & 0 \\
                   v\pi & -\frac{U}{2} & t_1 & 0  \\
                   0 & t_1 & \frac{U}{2} & v\pi^+ \\
                   0 & 0 & v\pi & \frac{U}{2}
                   \end{bmatrix}
 \end{gather}
We compute the eigenvectors corresponding to the eigenvalues $\epsilon_{\mathbf{k}\gamma}$ numerically.

\subsection{The screened interaction and self-energy} 
In the superfluid state the effective electron-hole interaction is given by\cite{lozovik2012,sodemann2012,Conti2019}
\begin{align}
W^{eh} = \frac{-V_D - \Pi_{a}(V^2 - V_D^2)}{1-2(V\Pi_{n} +V_D\Pi_{a}) +(\Pi_{n}^2 - \Pi_{a}^2)(V^2-V_D^2)}, 
\label{eq:W}
\end{align}
where the $k$ and Matsubara indexes are implicit for all quantities and
\begin{align}
&V=V_k=\frac{2\pi e^2}{\epsilon}\frac{1}{|\mathbf{k}|} \\
&V_D=V^D_k = e^{-dk} \frac{2\pi e^2}{\epsilon}\frac{1}{|\mathbf{k}|}. 
\end{align}
$\epsilon$ is the effective dielectric constant, $d$ the separation between the bilayers and $e$ the electronic charge.
We will now specialize to the case with equal electron hole dispersions so that $\epsilon_{1m}=\epsilon_{2m}$ and the electron and hole polarizations are equal. For this case the normal and anomoulous polarizations are given by 
\begin{align}
\Pi^{(n)}_0(q,i\nu_n) &=  4\sum_{\mathbf{k},\gamma,\gamma'} \mathcal{F}_{\mathbf{k},\mathbf{q}-\mathbf{k}}^{\gamma \gamma'} \left [ \frac{|u_{\mathbf{q}-\mathbf{k},\gamma'}|^2|u_{\mathbf{k}\gamma}|^2(n_F(\epsilon_{ \mathbf{k}\gamma})-n_F(\epsilon_{ \mathbf{q}-\mathbf{k}, \gamma'}))}{i\nu_n + \epsilon_{\mathbf{k}\gamma}-\epsilon_{\mathbf{q}-\mathbf{k}, \gamma'}} 
+ \frac{|u_{\mathbf{q}-\mathbf{k},\gamma'}|^2|v_{\mathbf{k}\gamma}|^2(n_F(-\epsilon_{\mathbf{k}\gamma})-n_F(\epsilon_{\mathbf{q}-\mathbf{k},\gamma'}))}{i\nu_n - \epsilon_{\mathbf{q}-\mathbf{k},\gamma'}-\epsilon_{\mathbf{k},\gamma}} \right. \nonumber \\
&+ \left. \frac{|u_{\mathbf{k},\gamma}|^2|v_{\mathbf{q}-\mathbf{k},\gamma'}|^2(n_F(\epsilon_{\mathbf{k}\gamma})-n_F(-\epsilon_{\mathbf{q}-\mathbf{k},\gamma'}))}{i\nu_n + \epsilon_{\mathbf{k}\gamma} + \epsilon_{\mathbf{q}-\mathbf{k},\gamma'}} 
+ \frac{|v_{\mathbf{q}-\mathbf{k},\gamma'}|^2|v_{\mathbf{k},\gamma}|^2(n_F(-\epsilon_{\mathbf{k},\gamma})-n_F(-\epsilon^-_{\mathbf{q}-\mathbf{k},\gamma'}))}{i\nu_n - \epsilon_{\mathbf{k}\gamma} + \epsilon_{\mathbf{q}-\mathbf{k},\gamma'}} \right ]
\end{align}

\begin{align}
 \Pi^{(a)}_0(q,i\nu_n) &=
 4\sum_{\mathbf{k},\gamma,\gamma'}\mathcal{F}_{\mathbf{k},\mathbf{q}-\mathbf{k}}^{\gamma \gamma'} \times \\ &\left [
 -\frac{u_{\mathbf{q}-\mathbf{k},\gamma'}v_{\mathbf{q}-\mathbf{k},\gamma'}^*u_{\mathbf{k}\gamma}^*v_{\mathbf{k}\gamma}(n_F(\epsilon_{\mathbf{q}-\mathbf{k},\gamma'})-n_F(\epsilon_{\mathbf{k}\gamma}))}{i\nu_n - \epsilon_{\mathbf{q}-\mathbf{k},\gamma'} + \epsilon_{\mathbf{k}\gamma}} +
 \frac{u_{\mathbf{q}-\mathbf{k},\gamma'}v_{\mathbf{q}-\mathbf{k},\gamma'}^*u_{\mathbf{k}\gamma}^*v_{\mathbf{k}\gamma}(n_F(\epsilon_{\mathbf{q}-\mathbf{k},\gamma'})-n_F(-\epsilon_{\mathbf{k}\gamma}))}{i\nu_n - \epsilon^+_{\mathbf{q}-\mathbf{k},\gamma'} - \epsilon^+_{\mathbf{k}\gamma}} \right. \nonumber \\
 &+ \left. \frac{u_{\mathbf{q}-\mathbf{k},\gamma'}v_{\mathbf{q}-\mathbf{k},\gamma'}^*u_{\mathbf{k}\gamma}^*v_{\mathbf{k},\gamma}(n_F(-\epsilon_{\mathbf{q}-\mathbf{k},\gamma'})-n_F(\epsilon_{\mathbf{k}\gamma}))}{i\nu_n + \epsilon_{\mathbf{q}-\mathbf{k},\gamma'} + \epsilon_{\mathbf{k}\gamma}}
 -\frac{u_{\mathbf{q}-\mathbf{k},\gamma'}v_{\mathbf{q}-\mathbf{k},\gamma'}^*u_{\mathbf{k}\gamma}^*v_{\mathbf{k}\gamma}(n_F(-\epsilon_{\mathbf{q}-\mathbf{k},\gamma'})-n_F(-\epsilon^+_{\mathbf{k}\gamma}))}{i\nu_n + \epsilon^-_{\mathbf{q}-\mathbf{k},\gamma'} - \epsilon^+_{\mathbf{k}\gamma}} \right]
\end{align}
Here the factor $4$ comes from the spin and valley degeneracy. 

In order to evaluate the frequency sums in the gap equation analytically we need an analytic expression for the self-energy. Therefore we have to fit $W$ to a rational function.
We choose the method of Ref. \onlinecite{Akashi13development} and fit $W$ to a plasmon pole function
\begin{align}
 \tilde{W}_{k}(i\nu_m)=W_{k}(0) + \sum_i^{N_p} a_{i;k}\left(\frac{2}{\omega_{i;k}}-\frac{2\omega_{i;k}}{\nu_m^2+\omega_{i;k}^2}\right)
\end{align}
where the $\omega_{ik}$ are the position of the poles which are obtained by a least-square fitting procedure as described below. In practice we compute $W$ along the imaginary axis using eq. \ref{eq:W} and continue it to the 
real axis using Pad\'e approximant. In order to be able to probe low temperatures we use an effective frequency mesh for the fitting procedure of $W(i\nu_m)$ which is sparser than the actual frequency mesh corresponding to the temperature of the calculations. However, all quantities (including the Fermi-Dirac distributions in the polarizations and the final frequency sums) are computed for the correct temperature. Therefore this approximation corresponds to an effective temperature smearing in the analytic continuation of $W$ and for the least-square fit. The resulting parameters $a_{i;nkn'k'}$ and $\omega_{i;nkn'k'}$ are not sensitive to the choice of mesh.     

\begin{figure}[tb]
\begin{center}
\includegraphics[width=0.85\textwidth]{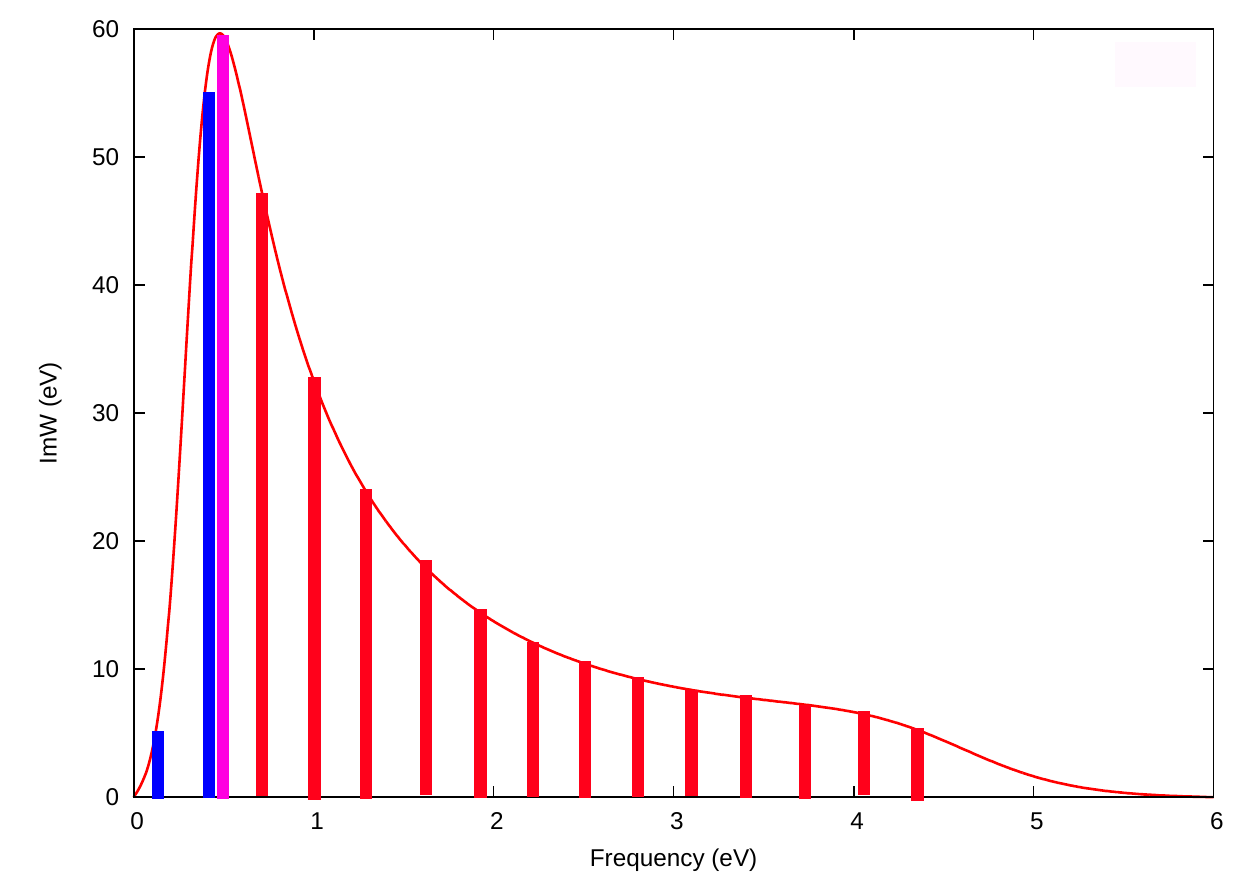} 
\caption{Sketch of the fitting procedure for $W$. The relative weight between the peaks within each train (the blue and red trains) are given by the strength of Im$W$, while the overall weight of each train is determined by a least-square fit.}
\label{fig:Wsketch} 
\end{center}
\end{figure}

For the bilayer bandstructure considered in this work Im$W_\mathbf{k}(\omega)$ only contains a single peak (Fig. \ref{fig:Wsketch}). However, since this peak is asymmetric with respect to the peak maximum it is not well represented by a single plasmon pole approximation. In order to remedy this we approximate $W$ by a multiple plasmon pole approximation with up to 100 poles. We use two trains with equidistant poles, one starting at some lowest frequency $\omega_\text{min}$ and ending just before the plasma frequency and the other starting at or after the plasma frequency and ending at the frequency $\omega_{\text{max}}$. $\omega_{\text{min}}$ and $\omega_{\text{max}}$ are determined as the frequencies $\omega'$ where max(Im$W(\omega)$)/Im$W(\omega') = \frac{1}{30}$  The relative weight between the peaks within each train is determined by the strength of Im$W$, while the overall weight of each train is determined by the least square fit. We also have an additional peak at the plasma frequency which is given a separate weight in the least square fit. Thus, we determine three parameters in the least square fit of $\tilde{W}(i\nu_m)$. The procedure is sketched in Fig. \ref{fig:Wsketch}.    

Then the self-energy is given by: 
\begin{align}
\Sigma^{12}_\gamma(\mathbf{q},i\omega_n) = \sum_{\mathbf{k},\gamma',\nu_l}\mathcal{F}_{\mathbf{q},\mathbf{q}-\mathbf{k}}^{\gamma,\gamma'}F^{eh}_{\gamma'}(\mathbf{q}-\mathbf{k},i\omega_n-i\nu_l)W^{eh}(\mathbf{k},\nu_l)=  \sum_{\mathbf{k},\gamma',\nu_l}\mathcal{F}_{\mathbf{q},\mathbf{q}-\mathbf{k}}^{\gamma,\gamma'}(\alpha_{\mathbf{k}\gamma'l} - \beta_{\mathbf{k}\gamma'l})
\end{align}

\begin{align}
\sum_{\mathbf{k},\gamma',\nu_l}\alpha_{\mathbf{k}\gamma'l} = \sum_{\mathbf{k},\gamma',\nu_l} \frac{v_{\mathbf{q}-\mathbf{k}}^{*}u_{\mathbf{q}-\mathbf{k}}}{i\omega_n-i\nu_l - \epsilon_{\gamma',\mathbf{q}-\mathbf{k}}} \left(W_{\mathbf{k}}(0) + \sum_i^{N_p} a_{i;\mathbf{k}}\left(\frac{2}{\omega_{i;\mathbf{k}}}-\frac{2\omega_{i;\mathbf{k}}}{\nu_l^2+\omega_{i;\mathbf{k}}^2}\right)\right) \nonumber \\
= -\sum_{\mathbf{k}\gamma'} \left[W_\mathbf{k}(0)+ \left( \sum_i^{N_p} a_{i;\mathbf{k}}\frac{2}{\omega_{i;\mathbf{k}}} \right)\right] v^*_{\mathbf{q}-\mathbf{k}}u_{\mathbf{q}-\mathbf{k}}n_F(-\epsilon_{\gamma',\mathbf{q}-\mathbf{k}}) 
- \sum_{\mathbf{k},i} v^*_{\mathbf{q}-\mathbf{k}}u_{\mathbf{q}-\mathbf{k}} \frac{2a_{i\mathbf{k}}\omega_{i\mathbf{k}} n_F(-\epsilon^+_{\mathbf{q}-\mathbf{k}})}{\omega_{i\mathbf{k}}^2 + \omega_n^2+2i\omega_n\epsilon^+_{\mathbf{q}-\mathbf{k}}-(\epsilon^+_{\mathbf{q}-\mathbf{k}})^2)} \nonumber \\
+ \frac{a_{i\mathbf{k}}n_B(-\omega_{i\mathbf{k}})}{\omega_{i\mathbf{k}} + i\omega_n - \epsilon^+_{\mathbf{q}-\mathbf{k}}} + \frac{a_{i\mathbf{k}}n_B(\omega_{i\mathbf{k}})}{\omega_{i\mathbf{k}} - i\omega_n + \epsilon^+_{\mathbf{q}-\mathbf{k}}}
\end{align}

\begin{align}
\sum_{\mathbf{k},\nu_l}\beta_{\mathbf{k}l} = \sum_{\mathbf{k},\nu_l} \frac{v_{\mathbf{q}-\mathbf{k}}^{*}u_{\mathbf{q}-\mathbf{k}}}{i\omega_n-i\nu_l + \epsilon_{\mathbf{q}-\mathbf{k}}^-} \left(W_{\mathbf{k}}(0) + \sum_i^{N_p} a_{i;\mathbf{k}}\left(\frac{2}{\omega_{i;\mathbf{k}}}-\frac{2\omega_{i;\mathbf{k}}}{\nu_l^2+\omega_{i;\mathbf{k}}^2}\right)\right) \nonumber \\
= -\sum_\mathbf{k} \left[W_\mathbf{k}(0)+ \left( \sum_i^{N_p} a_{i;\mathbf{k}}\frac{2}{\omega_{i;\mathbf{k}}} \right)\right] v^*_{\mathbf{q}-\mathbf{k}}u_{\mathbf{q}-\mathbf{k}}n_F(\epsilon^-_{\mathbf{q}-\mathbf{k}}) 
- \sum_{\mathbf{k},i} v^*_{\mathbf{q}-\mathbf{k}}u_{\mathbf{q}-\mathbf{k}} \frac{2a_{i\mathbf{k}}\omega_{i\mathbf{k}} n_F(\epsilon^-_{\mathbf{q}-\mathbf{k}})}{\omega_{i\mathbf{k}}^2 + \omega_n^2-2i\omega_n\epsilon^-_{\mathbf{q}-\mathbf{k}}-(\epsilon^-_{\mathbf{q}-\mathbf{k}})^2)} \nonumber \\
+ \frac{a_{i\mathbf{k}}n_B(-\omega_{i\mathbf{k}})}{\omega_{i\mathbf{k}} + i\omega_n + \epsilon^-_{\mathbf{q}-\mathbf{k}}} + \frac{a_{i\mathbf{k}}n_B(\omega_{i\mathbf{k}})}{\omega_{i\mathbf{k}} - i\omega_n - \epsilon^-_{\mathbf{q}-\mathbf{k}}}
\end{align}
where $n_F$ is the Fermi-dirac distribution and $n_B$ the Bose-Einstein distribution.
The self-energies contain four different kinds of terms:
\begin{align}
 \Sigma_0&=c_0\\
 \Sigma_1(i\omega_n)&=\frac{c_1}{\omega_n^2 -f_1 i\omega_n + d_1}\\
 \Sigma_2(i\omega_n)&=\frac{c_2}{d_2-i\omega_n}\\
 \Sigma_3(i\omega_n)&=\frac{c_3}{d_3+i\omega_n}
\end{align}

These expressions can be inserted in the gap equation and the Matsubara sums evaluated analytically using the MatsubaraSum Mathematica package \cite{MatsubaraSum}. 

\subsection{k-point sampling}
We use circular coordinates for the k-point sampling.
In the radial $\mathbf{k}$-direction we use an exponential mesh, which is densest around $\mathbf{k}=0$ and $|\mathbf{k}|=k_F$ where $k_F = \sqrt{\pi n}$. 
For $|\mathbf{k}|<k_F/2$ it is given by
\begin{align*}
k_i=C*(exp(x_i)-1) \\
x_i=(i-1)*\alpha_1/(N^a_k-1) \\
C=k_{F}/(2\cdot exp(\alpha_1)-1).
\end{align*}
The mesh is then mirrored to $k_F$ and mirrored again to $3k_F/2$. After $3k_F/2$ the mesh is given by
\begin{align*}
k_i=\frac{3k_F}{2} + C*(exp(x_i)-1) \\
x_i=(i-N^a_{k})*\alpha_1/(N_k-N_k^a) \\
C=k_{cut}-1.5\cdot k_F/( exp(\alpha_2)-1).
\end{align*}
where $N_k^a$ are the number of k-points below $\frac{3k_F}{2}$ and $N_k$ the total number of k-points. We use 150-200 k-points with $N_k^a \approx \frac{3}{4}N_k$, $\alpha_1 \approx 3.5$ and $\alpha_2 \approx 5.5$. For the angular part $\theta$ we use an equidistant mesh with 50-100 points.

\bibliography{refs}